\documentclass{aastex}
\usepackage{spr-astr-addons}
\usepackage{url}

\DeclareMathAlphabet{\mathbi}{\encodingdefault}{\rmdefault}{\bfdefault}{\itdefault}
\renewcommand{\vec}[1]{\ensuremath{\mathbi{#1}}}

\renewcommand{\eqref}[1]{Equation (\ref{#1})}

\def\lapp{\ifmmode\stackrel{<}{_{\sim}}\else$\stackrel{<}{_{\sim}}$\fi}
\def\gapp{\ifmmode\stackrel{>}{_{\sim}}\else$\stackrel{<}{_{\sim}}$\fi}

\newcommand{\emaila}{yanwm@uao.ac.cn}

\begin{document}

\title{Rotation measure variations for 20 millisecond pulsars}

\author{W. M. Yan\altaffilmark{1,2,3}}
\and \author{R. N. Manchester\altaffilmark{3}}
\and \author{G. Hobbs\altaffilmark{3}}
\and \author{W. van Straten\altaffilmark{4}} \and \author{J. E. Reynolds\altaffilmark{3}}
\and \author{N. Wang\altaffilmark{1,5}} \and \author{M. Bailes\altaffilmark{4}} \and \author{N. D. R. Bhat\altaffilmark{4}}
\and \author{S. Burke-Spolaor\altaffilmark{3}} 
\and \author{D. J. Champion\altaffilmark{6}} \and \author{A. Chaudhary\altaffilmark{3}}
\and \author{W. A. Coles\altaffilmark{7}} \and \author{A. W. Hotan\altaffilmark{3}}
\and \author{J. Khoo\altaffilmark{3}} \and \author{S. Oslowski\altaffilmark{4,3}}
\and \author{J. M. Sarkissian\altaffilmark{3}} \and \author{D. R. B. Yardley\altaffilmark{8,3}}

\altaffiltext{1}{Xinjiang Astronomical Observatory, Chinese Academy of
     Sciences, 150 Science-1 Street, Urumqi, Xinjiang 830011, China \emaila}
\altaffiltext{2}{Graduate University of Chinese Academy of Sciences, 19A Yuquan
  Road, Beijing 100049, China}
\altaffiltext{3}{CSIRO Astronomy and Space Science, Australia Telescope
  National Facility, PO Box 76, Epping NSW 1710, Australia}
\altaffiltext{4}{Centre for Astrophysics and Supercomputing, Swinburne University of 
Technology, PO Box 218, Hawthorn VIC 3122, Australia}
\altaffiltext{5}{Key Laboratory of Radio Astronomy, Chinese Academy of Sciences， 
Nanjing, 210008, China}
\altaffiltext{6}{Max-Planck-Institut f\"{u}r Radioastronomie, Auf dem H\"{u}gel 69, 
53121 Bonn, Germany}
\altaffiltext{7}{Electrical and Computer Engineering, University of California at
  San Diego, La Jolla, California, USA}
\altaffiltext{8}{Sydney Institute for Astronomy, School of Physics, The University 
of Sydney, NSW 2006, Australia}

\begin{abstract}
  We report on variations in the mean position angle of the 20
  millisecond pulsars being observed as part of the Parkes Pulsar
  Timing Array (PPTA) project. It is found that the observed
  variations are dominated by changes in the Faraday rotation
  occurring in the Earth's ionosphere. Two ionospheric models are used
  to correct for the ionospheric contribution and it is found that one
  based on the International Reference Ionosphere gave the best
  results. Little or no significant long-term variation in
  interstellar RM was found with limits typically about 0.1 rad
  m$^{-2}$ yr$^{-1}$ in absolute value. In a few cases, apparently
  significant RM variations over timescales of a few 100 days or more
  were seen. These are unlikely to be due to localised magnetised regions
  crossing the line of sight since the implied magnetic fields are too
  high. Most probably they are statistical fluctuations due to random
  spatial and temporal variations in the interstellar electron density
  and magnetic field along the line of sight. 
\end{abstract}


\section{Introduction}
\label{sect:intro}

The observed radiation from pulsars is affected by propagation effects
along the path between the pulsar and the observatory. One of the
various propagation effects is Faraday rotation, that is, the rotation
of the plane of linear polarization caused by the different phase
velocities of the two hands of circular polarization in a magnetised
plasma. Pulsar radiation typically has strong
linear polarization making it relatively easy to observe Faraday
rotation for these sources.  Faraday rotation is described by the rotation
measure (RM), defined by
\begin{equation}\label{eq:fara}
  \psi = {\rm RM}\; \lambda^2,
\end{equation}
where $\psi$ is the linear polarization position angle
(PA) and $\lambda = c/\nu$ is the radio wavelength corresponding to radio
frequency $\nu$. The rotation measure is approximated by
\begin{equation}\label{eq:rm}
{\rm RM} = 0.810 \int_0^D n_e \vec{B} \cdot d\vec{l},
\end{equation}
where $n_e$ is the interstellar electron density in units of
cm$^{-3}$, $\vec{B}$ is the vector magnetic field in microgauss, 
$d\vec{l}$ is an elemental vector along the line-of-sight toward us
(positive RMs correspond to fields directed towards us) in
parsecs and RM is in units of rad m$^{-2}$. 

Along with the dispersion measure (DM), defined by 
\begin{equation}\label{eq:dm}
{\rm DM} = \int_0^D n_e\;dl,
\end{equation}
where DM is in units of cm$^{-3}$~pc, the RM may be used to determine
the strength and direction of the interstellar magnetic field
\citep[see, e.g.,][]{man74,hml+06,njkk08}. Combining Equations \ref{eq:rm}
and \ref{eq:dm}, we obtain an estimate of the mean line-of-sight
component of the interstellar magnetic field weighted by the local
electron density along the path:
\begin{equation}\label{eq:bpar}
\langle B_{||}\rangle = 1.232\; \frac{\rm RM}{\rm DM}\;\mu{\rm G},
\end{equation}
where RM and DM are in their usual units.

Long-term changes in pulsar PAs can result from changes in the
polarization of the emitted radiation or from changes in the RM along
the path (\eqref{eq:fara}). Intrinsic PA changes could result from
precession of the pulsar spin axis or changes in the magnetic
configuration of the star. RM changes can occur as the path to the
pulsar traverses different regions of the interstellar medium (ISM) or
in the Earth's ionosphere due to the diurnal and other changes in the
ionospheric total electron content.

Correlated RM and DM variations with time have been observed for
several pulsars.  \citet{hmm+77,hhc85} analyze the variations of RM
and DM for the Vela pulsar, finding an approximately linear variation
in both with slopes of 0.73~rad m$^{-2}$~yr$^{-1}$ and
$-0.04$~cm$^{-3}$pc~yr$^{-1}$ for RM and DM respectively. They
interpret the RM and DM variations as the result of the relative
motion of a magnetized filament of the Vela supernova remnant across
the line-of-sight and estimated the electron density and the effective
mean line-of-sight component of magnetic field in the
filament. Adopting 500 km s$^{-1}$ as the transverse velocity of the
source, they obtain a transverse scale length of $5.1\times 10^{-4}$
pc per year. Assuming that the line-of-sight dimension is comparable
with the transverse scale gives an electron density in the region of
$\Delta n_e = 78\; \rm cm^{-3}$. The mean line-of-sight component of
magnetic field in the varying region is given by 
\begin{equation}\label{eq:bvar}
\langle B_{||}\rangle_{\rm var}=1.23\;(d{\rm RM}/dt)$$/(d{\rm DM}/dt)\;\mu{\rm
  G}
\end{equation}
(cf. Equation~\ref{eq:bpar}) and for Vela is $22\mu$G.

\citet{vdhm97} made a similar analysis for PSRs B1556$-$44 and
B1727$-$47. For PSR B1556$-$44, the estimated electron density and the
effective mean line-of-sight component of magnetic field in the
varying region are $\Delta n_e = 290\; \rm cm^{-3}$ and $\langle
B_{||}\rangle_{\rm var} = 2\; \mu{\rm G}$ directed away from the pulsar; for PSR
B1727$-$47, the electron density is $\Delta n_e = 170\; \rm cm^{-3}$,
the magnetic field strength is $\langle B_{||}\rangle_{\rm var} = 16\; \mu{\rm
  G}$ directed toward the pulsar. Because of the absence of an
associated supernova remnant the authors proposed that the RM
variation for PSR B1727$-$47 may be due to a an irregularity within
the {\sc Hii} region of I Scorpius located at a distance
of 1.3 kpc.  \citet{rcip88} made a detailed analysis of the RM and DM
variations for the Crab pulsar and found correlated variations between
1972 and 1974 which imply a magnetic field in the varying region of
about 170~$\mu$G. Several other authors \citep[e.g.][]{wck+04,hml+06}
also find apparent time variations in the interstellar RMs of several
pulsars.

Much stronger fields are observed in vicinity of stars. For example,
\citet{jbwm05} observed RM variations of many 1000 rad m$^{-2}$ and DM
variations of $\sim 20$~cm$^{-3}$~pc for signals propagating through
the disk of SS 2883, the companion star to PSR B1259$-$63, implying
fields in the mG range. RM variations of a few 100 rad m$^{-2}$ have
been observed as the path to PSR J1801$-$2304 passed close to the Sun
\citep{ojs07}. With expected values of electron density in the solar
corona, these variations again imply fields of order 1~mG.

The Parkes Pulsar Timing Array (PPTA) project is making regular timing
observations of 20 millisecond pulsars (MSPs) with the primary goal of
detecting low-frequency (nHz) gravitational waves
\citep{man08,vbb+10}.  Data from the first three years of this project
(2004 -- 2007) were used by \citet{yhc+07} to investigate the DM
variations in the PPTA pulsars. Significant variations were found in
many of the pulsars, with observed rates of change up to $\sim
10^{-2}$~pc~cm$^{-3}$~yr$^{-1}$. The PPTA sample is unusual if not
unique in having a large number of observations over a long data span
with high signal/noise ratio and accurately calibrated polarization
data for a significant number of pulsars. Mean pulse total intensity
and polarization profiles have been published for the 20 PPTA pulsars
by \citet{ymv+11} showing that the pulsed emission extends over more
than half of the pulse period for most of the sample and that the
polarization variations across the profile are generally complex.

In this paper we report on the long-term variations of observed mean
PAs over a data span of 4.8 years for the 20 PPTA MSPs. It is shown
that, after correcting for Faraday rotation in the Earth's ionosphere,
there is little or no significant variation in the observed PAs and
hence also in the interstellar RMs. Details of the observations and data
processing are given in \S\ref{sect:obs}, a comparison between two different
ionospheric RM correction methods is made in \S\ref{sect:iono}, the observed
variations are discussed in \S\ref{sect:disc} and we provide a short summary in
\S\ref{sect:concl}.

\section{Observations and analysis}
\label{sect:obs}
The PPTA observations were made at intervals of 2 -- 3 weeks using the
Parkes 64-m radio telescope with the centre beam of the 20cm Multibeam
receiver \citep{swb+96}. The Multibeam receiver has a bandwith of
about 300 MHz centered at about 1.4 GHz. The system equivalent flux
density on cold sky is approximately 30 Jy. The Multibeam receiver has
an orthogonal linearly polarized feed system. There is a calibration
probe at 45\degr~to the two signal probes through which a linearly
polarized broadband and pulsed calibration signal can be
injected. Three different back-end systems were used, namely the Parkes digital
filterbank systems PDFB1, PDFB2 and PDFB4. The PDFB systems are
digital polyphase filterbanks using field-programmable gate array
(FPGA) processors. Signals from each of the two orthogonal
polarizations are band-limited to 256 MHz and 8-bit sampled at the
Nyquist rate. After transformation to the spectral domain, typically
with 1024 channels across the band, data are
folded at the apparent pulse period to form mean pulse profiles in the
four Stokes parameters.

All data were recorded using the PSRFITS data format \citep{hvm04}
with 1-minute sub-integrations. For PSRs J1857+0943, J1939+2134 and
J2124$-$3358, each observation was 32 minutes duration; for the other
17 pulsars the observation times were 64 minutes. A short 2-minute
observation of the pulsed calibration signal was made before each
pulsar observation. Flux density scales were established using
observations of Hydra A which was assumed to have a flux density of
43.1 Jy at 1400 MHz and a spectral index of $-0.91$. The Multibeam
receiver has significant coupling between the two signal feed probes. The
cross-coupling was measured by analysis of series of observations of
PSR J0437$-$4715 covering a wide range of parallactic angles
\citep{vs04}.

All observations were processed using the {\sc psrchive} software
package \citep{hvm04}. Strong narrow-band radio frequency interference
(RFI) was identified by searching for spectral channels that were
significantly above the local mean; such channels were excised from
the data. Similarly, broad-band impulsive RFI was identified on
zero-DM summed time-domain data and excised. The variations in
instrumental gain and phase across the band and the effects of
cross-couping in the feed were removed using the calibration
observations. Polarization parameters are in accordance with the
astronomical conventions described by \citet{vmjr10}; in particular,
PAs are absolute and referred to celestial north, increasing toward
east. Table \ref{tab:para} lists the basic parameters for the 20 PPTA
MSPs with RM values derived from PDFB2 observations following
\citet{ymv+11}.

\begin{table*}
\small
\centering
\begin{minipage}[]{120mm}
\caption[]{ Parameters for the 20 PPTA pulsars}\label{tab:para}\end{minipage}
\tabcolsep 6mm
 \begin{tabular}{cccccc}
  \hline\noalign{\smallskip}
PSR  & $P$ &DM  & RM  &Data span &Nr of \\
	    &(ms) &(cm$^{-3}$ pc)   &(rad m$^{-2}$)   &(MJD)  &Obs.\\
  \hline\noalign{\smallskip}
J0437$-$4715     &5.757 &2.64      & $-$0.6 &53703 -- 55312   &393   \\
J0613$-$0200     &3.062  &38.78    & 9.7 &53687 -- 55407  &125  \\
J0711$-$6830     &5.491  &18.41    &21.4  &53660 -- 55408  &47  \\
J1022$+$1001       &16.453 &10.25   & $-$0.3 &53727 -- 55406  &84  \\
J1024$-$0719     &5.162   &6.49       &$-$8.3 &53688 -- 55407  &79  \\\\
J1045$-$4509     &7.474   &58.17    &92.3 &53660 -- 55408  &110  \\
J1600$-$3053     &3.598  &52.33   &$-$15.8  &53660 -- 55406  &103  \\
J1603$-$7202    &14.842  &38.05    &27.7  &53703 -- 55406  &101  \\
J1643$-$1224    &4.622   &62.41    &$-$308.2  &53724 -- 55406  &98  \\
J1713$+$0747     &4.570    &15.99   &9.3  &53660 -- 55406  &112  \\\\
J1730$-$2304     &8.123  &9.62    &$-$7.2  &53725 -- 55387  &70  \\
J1732$-$5049   &5.313     & 56.82   &$-$6.7  &53799 -- 55117  & 22 \\
J1744$-$1134   &4.075    &3.14    & $-$1.6 &53660 -- 55406  &107  \\
J1824$-$2452    &3.054   &120.50    &77.6  &53688 -- 55350  &76  \\
J1857$+$0943       &5.362  &13.30   &18.2  &53703 -- 55406  &62  \\\\
J1909$-$3744    & 2.947  &10.39    &$-$6.5 &53688 -- 55406  &174  \\
J1939$+$2134       & 1.558 &71.04   &6.9   &53816 -- 55406  &76  \\
J2124$-$3358     & 4.931 &4.60     &$-$5.4  &53660 -- 55406  &103  \\
J2129$-$5721     & 3.726 &31.85    &23.5  &53724 -- 55389  &62  \\
J2145$-$0750     &16.052 &9.00    & $-$1.3 &53725 -- 55389  &67  \\
  \noalign{\smallskip}\hline
\end{tabular}
\end{table*}

To measure the long-term changes in mean PA, we took the weighted
average of the differences between the PAs across an individual
observation profile and the PAs across the grand-average profile from
\citet{ymv+11}. This gives a relative PA shift for that observation
and its estimated uncertainty. This is then repeated for all
observations of a given pulsar to give the PA variations across the
data span. In order to accurately align the observation profiles for a
given pulsar, pulse times of arrival were obtained for all the
observations and fitted using the TEMPO2 pulsar timing package
\citep{hem06} to give a timing model for the pulsar. This timing model
was then used to align the profiles before computing the PA
differences.

Because the Earth's magnetic field is relatively strong ($\sim 0.5$~G)
and the electron density in the Earth's ionosphere relatively high (up
to $10^6$~cm$^{-3}$), the Earth's ionosphere makes a signficant
contribution to the total RM along the path to the pulsar. For the
Southern Hemisphere, the Earth's magnetic field is directed outwards and so the
ionospheric contribution to the RM is negative, typically between
$-0.5$ and $-3$ rad m$^{-2}$, with a strong diurnal variation due to
the varying ionspheric electron density. 

\section{Ionospheric RM corrections}
\label{sect:iono}
We have used two programs to compute the ionospheric RM (RMi) in the
direction of a given source at a given time. The first, {\sc farrot},
was developed at the Dominion Radio Astrophysical Observatory (DRAO),
Penticton, Canada. It uses three Chapman layers to model the D, E and
F regions and a semi-empirical model for variations in these based on the
10.7-cm solar radio flux measured by DRAO. The second, {\sc
  getRM\_iono}, was originally written by J. L. Han for correction of
measured pulsar RMs \citep[e.g.,][]{hml+06}. In its current form, it
uses the 2007 International Reference Ionosphere (IRI)
model\footnote{See http://iri.gsfc.nasa.gov for a general description
  of the IRI. Source and data files are available from
  http://nssdcftp.gsfc.nasa.gov/models/ionospheric/iri/iri2007/.} with
the URSI model for the critical frequency of the F2 ionospheric layer
($f_0$F2) and input of the geomagnetic Ap index, the solar 10.7-cm
radio flux and sunspot numbers. Both models use the International
Geomagnetic Reference Field\footnote{See
  http://www.ngdc.noaa.gov/IAGA/vmod/igrf.html} to define the
structure of the Earth's magnetic field.

\begin{figure*}
\centering
\includegraphics[angle=-90,scale=.328]{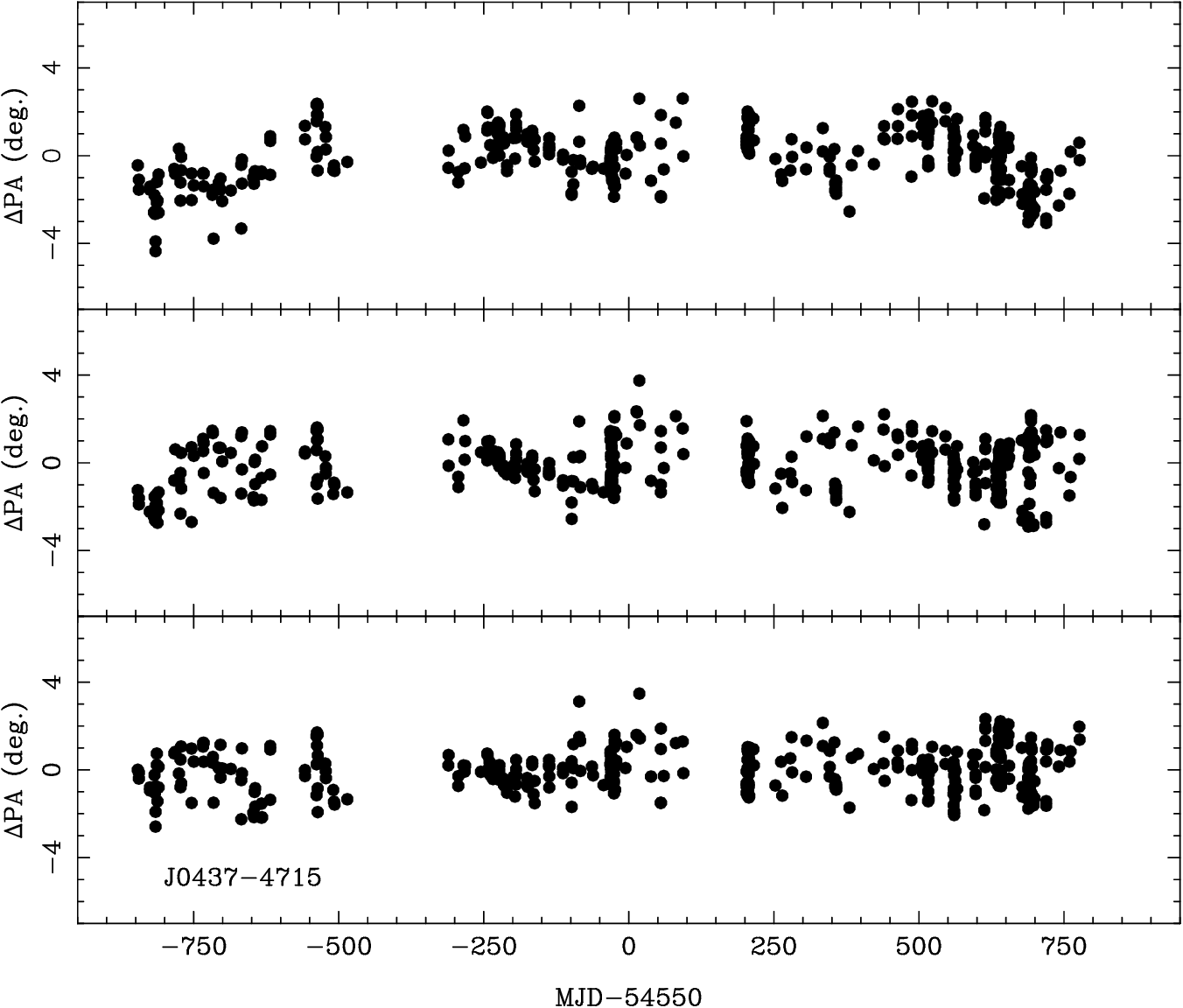}
\includegraphics[angle=-90,scale=.328]{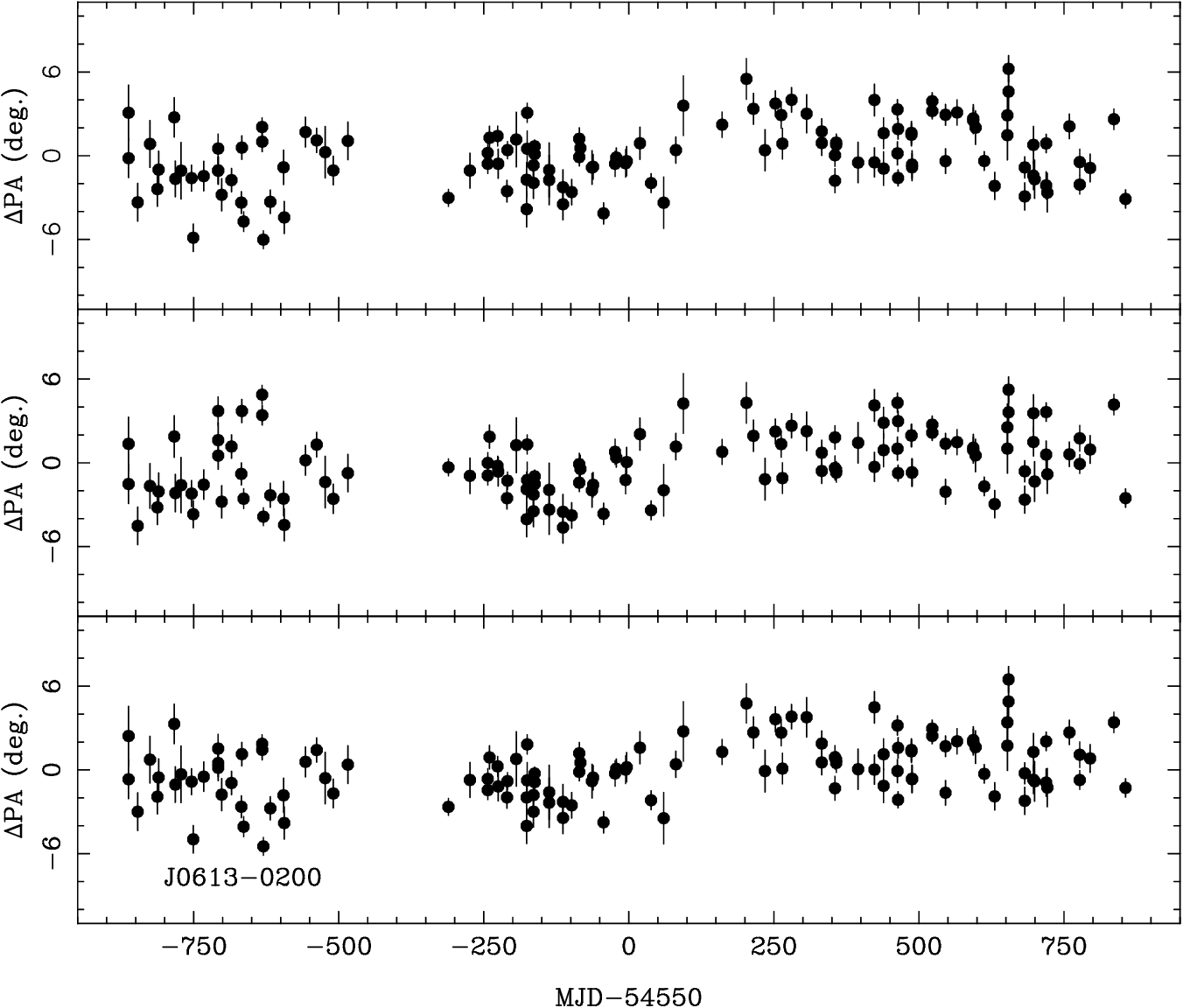}
\includegraphics[angle=-90,scale=.328]{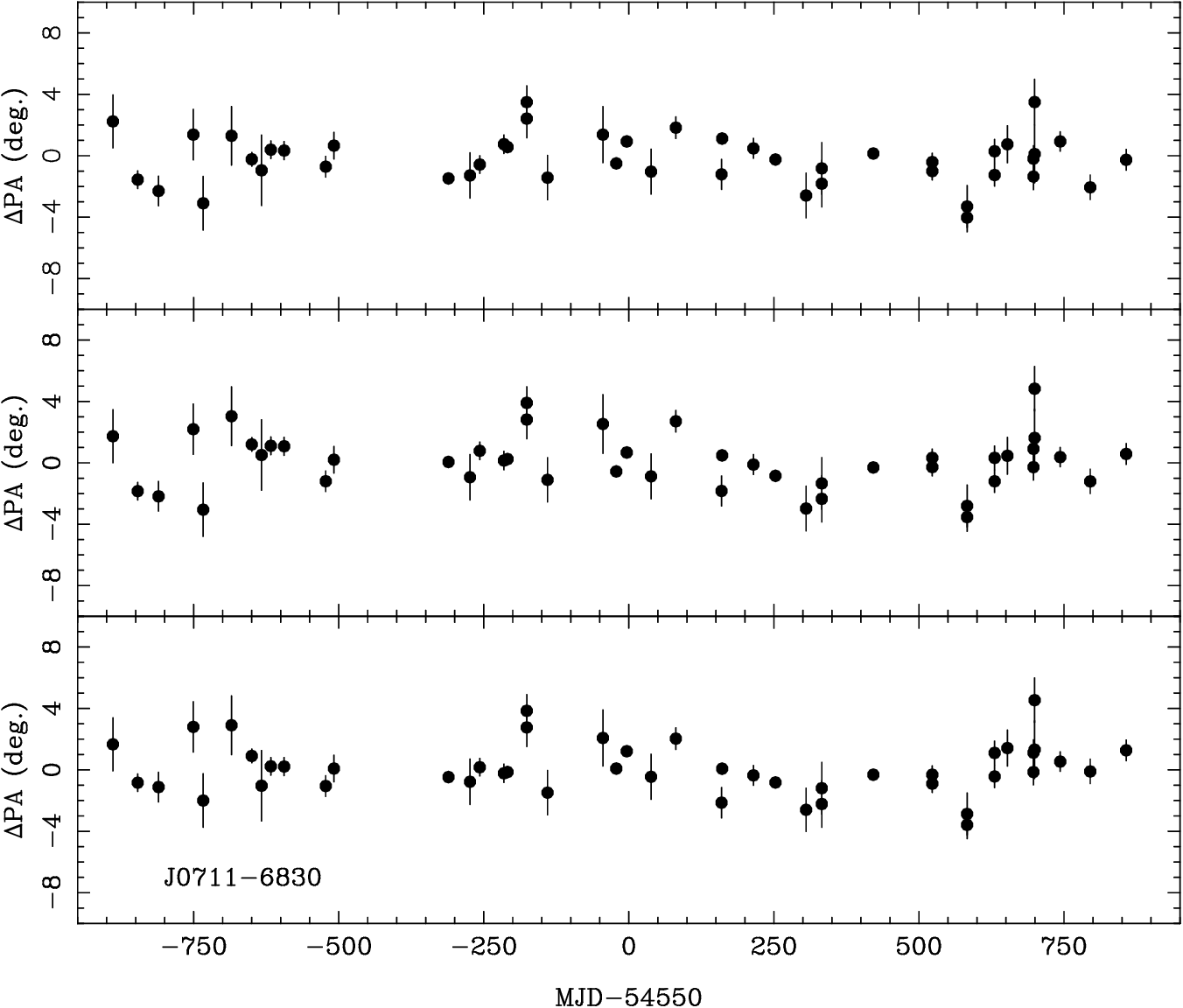}
\includegraphics[angle=-90,scale=.328]{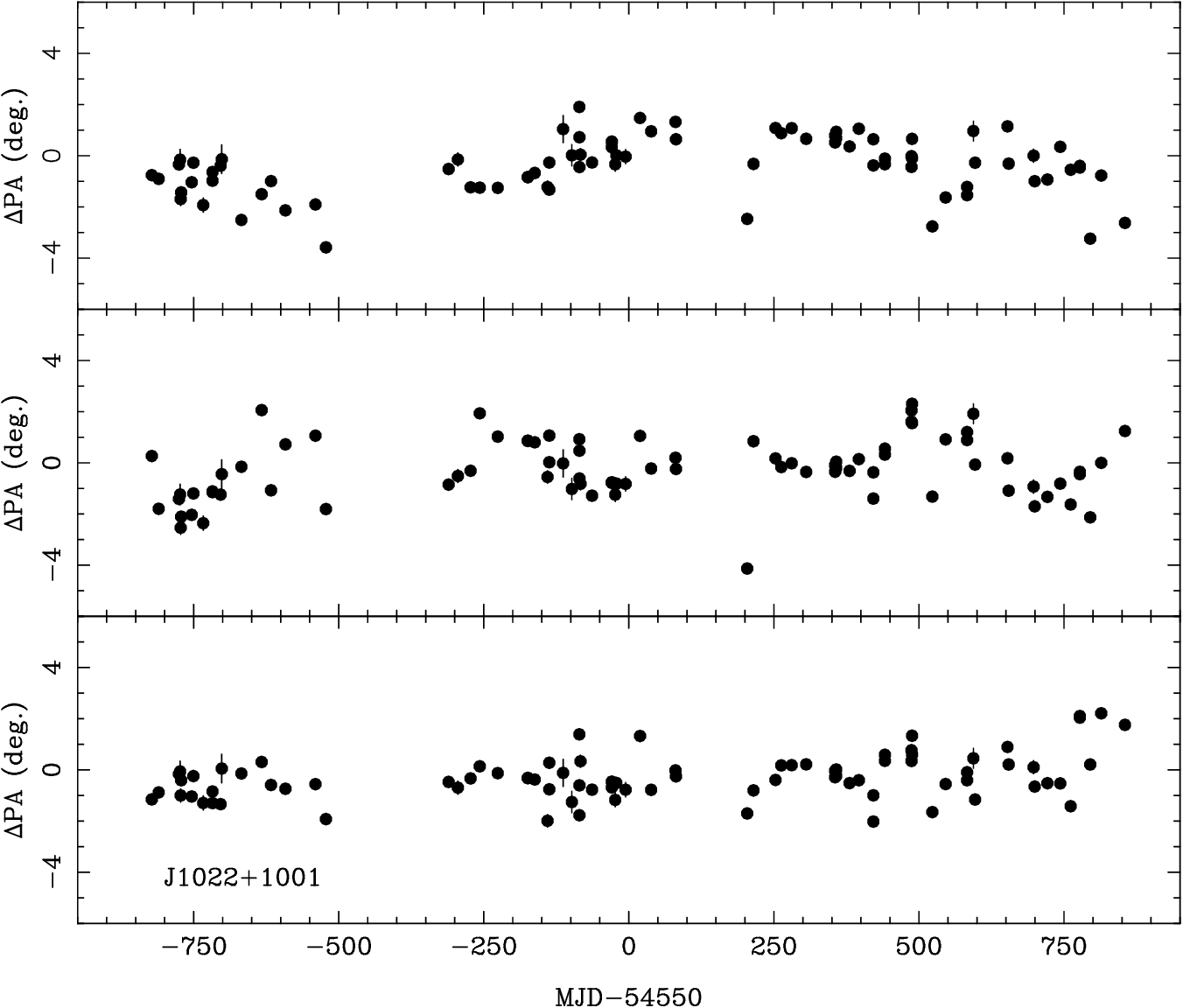}
\includegraphics[angle=-90,scale=.328]{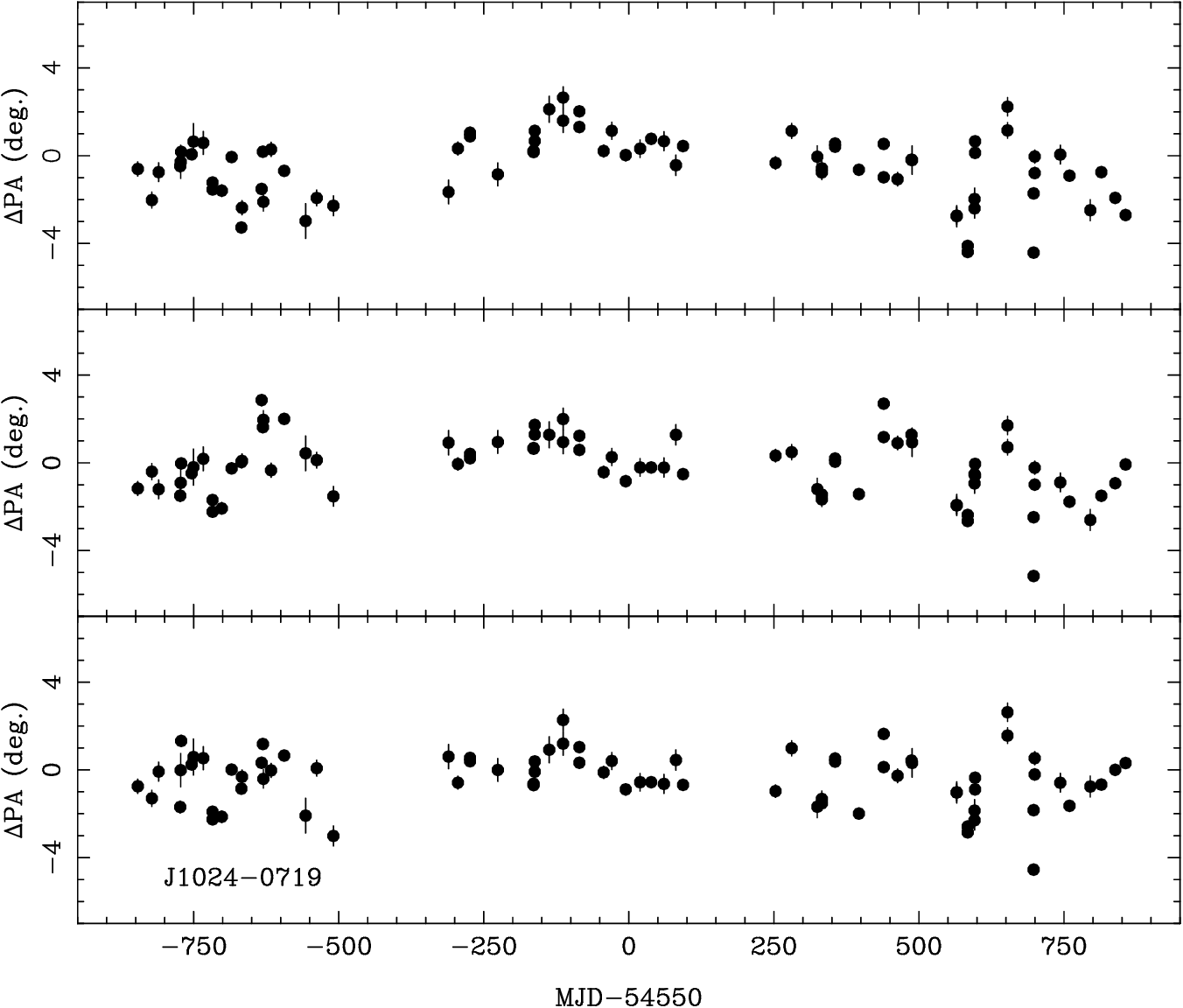}
\includegraphics[angle=-90,scale=.328]{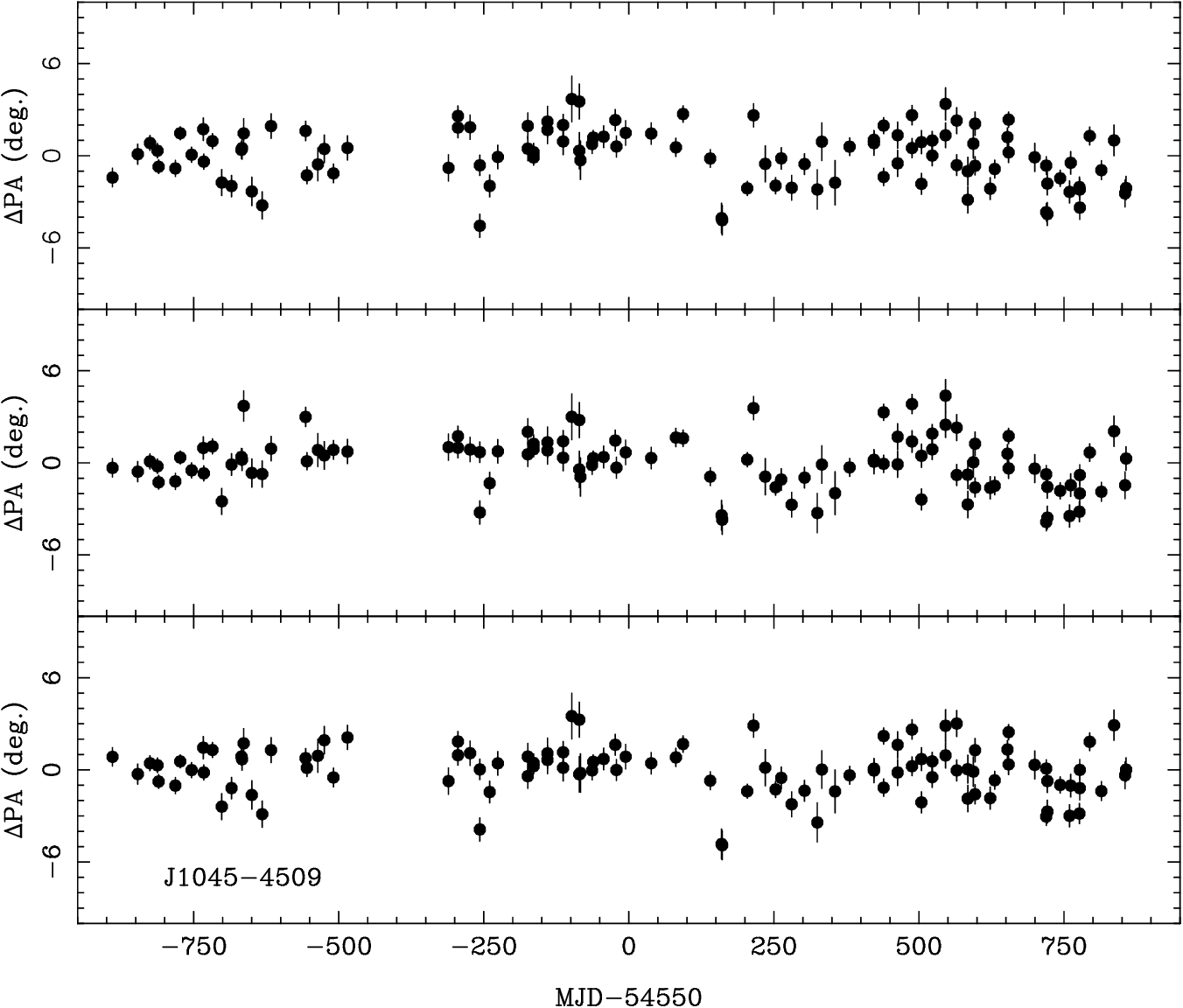}
\caption{PA differences as a function of time for six of the PPTA
  pulsars. In each sub-figure, the upper panel shows PA differences
  without ionospheric correction, the middle panel gives PA
  differences after ionospheric corrections using the Penticton
  program {\sc farrot}; the lower part shows differences after
  ionospheric corrections using the IRI ionospheric
  model.\label{Fig:padiff1}}
\end{figure*}

\begin{figure*}
\centering
\includegraphics[angle=-90,scale=.328]{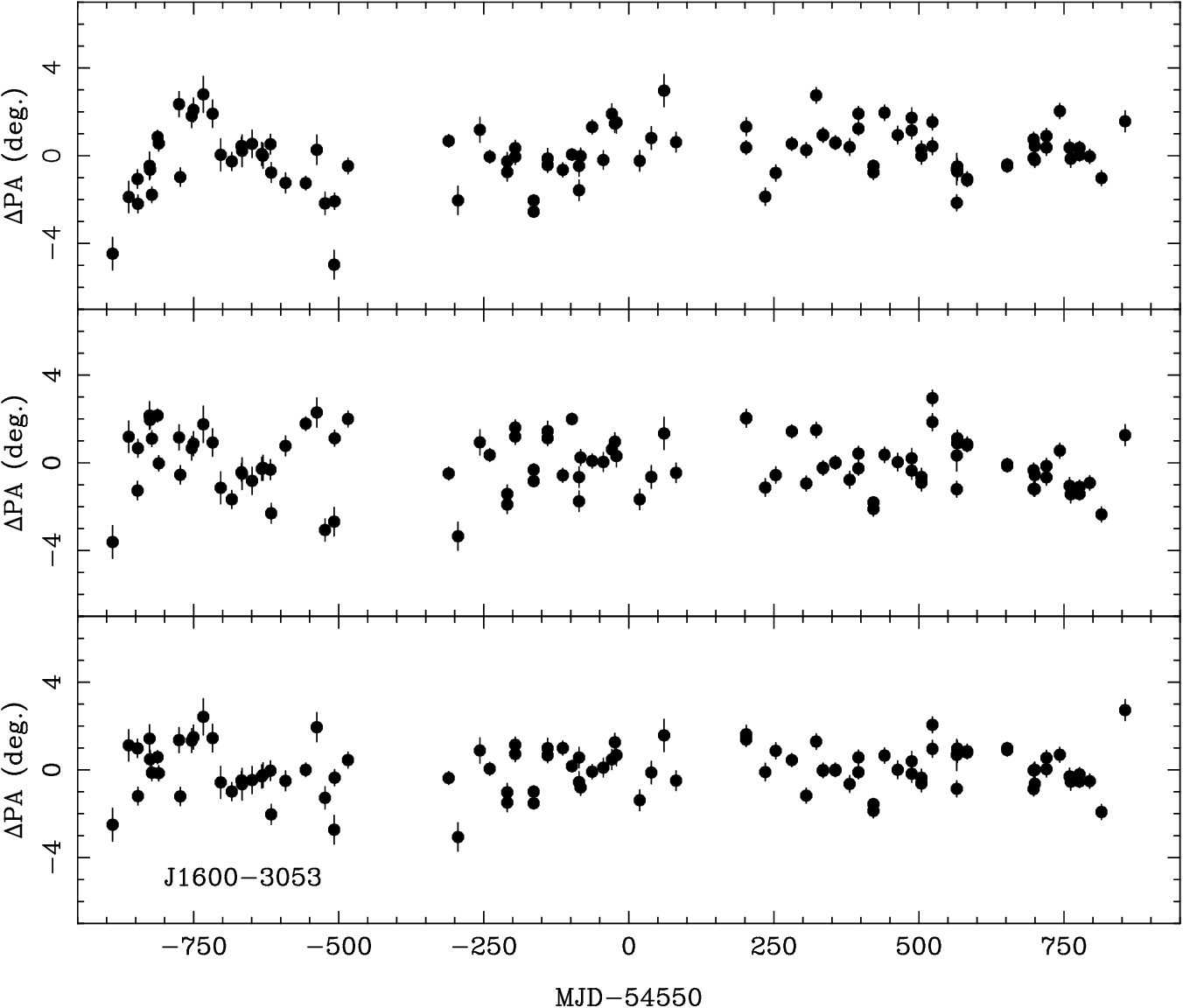}
\includegraphics[angle=-90,scale=.328]{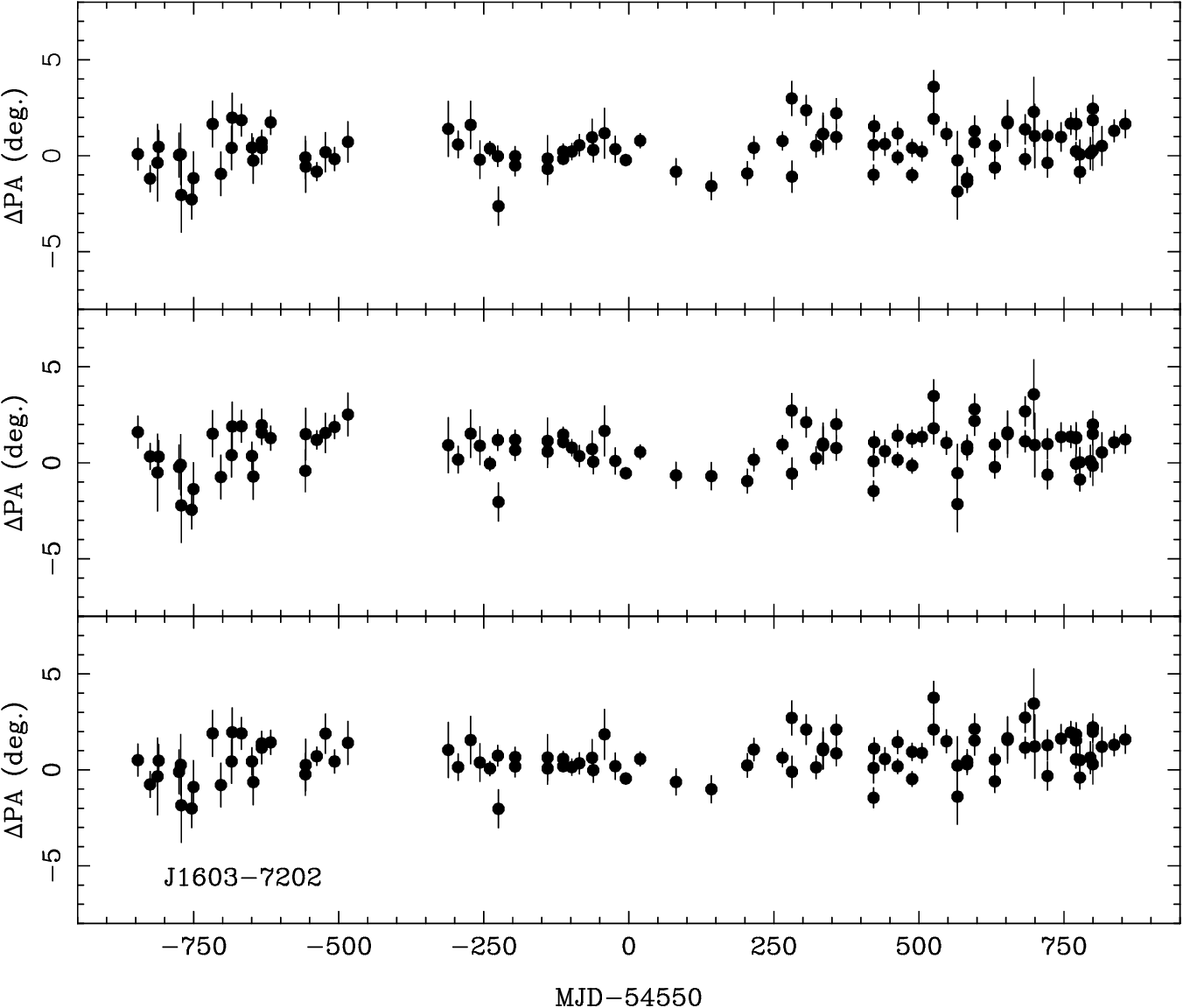}
\includegraphics[angle=-90,scale=.328]{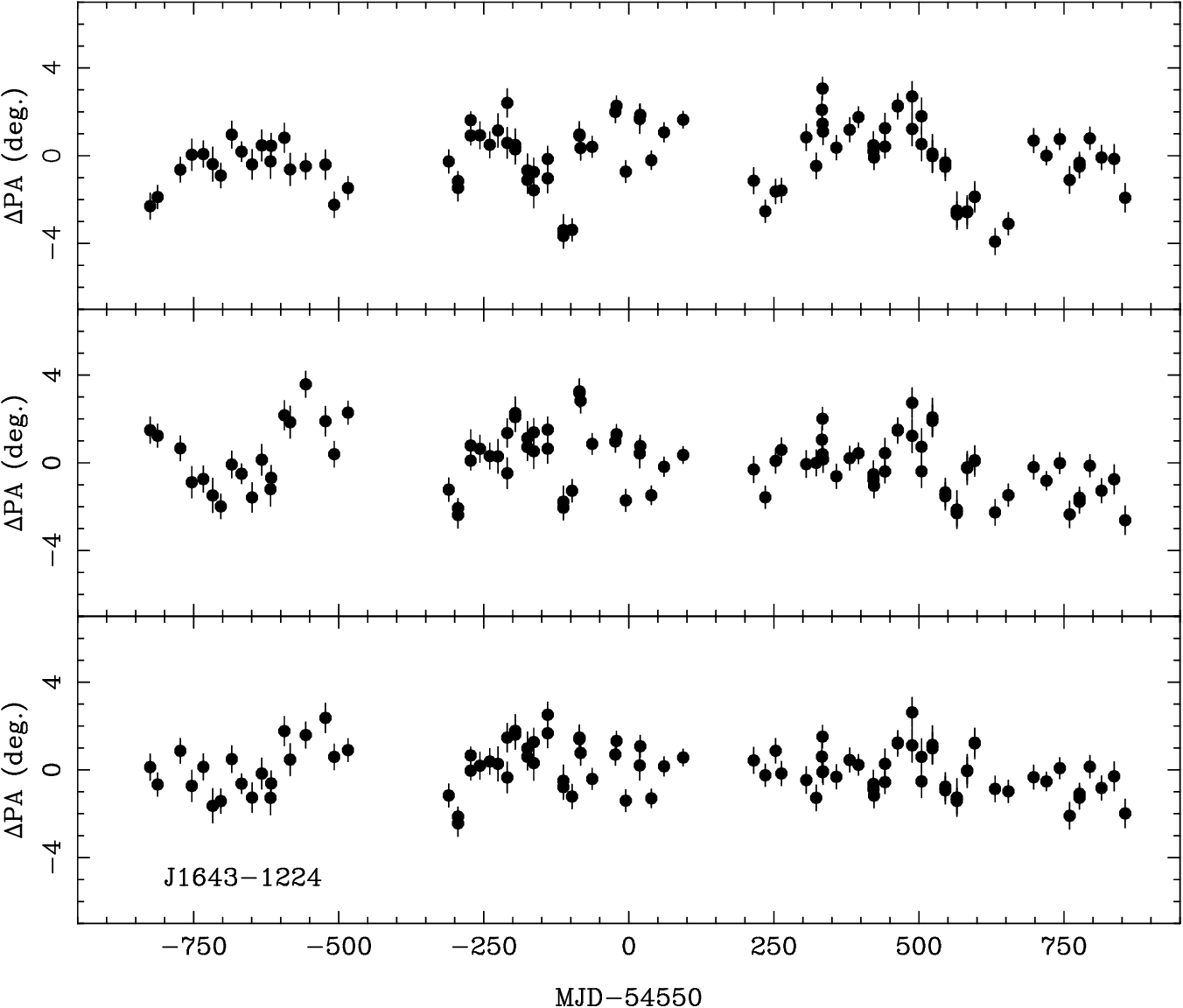}
\includegraphics[angle=-90,scale=.328]{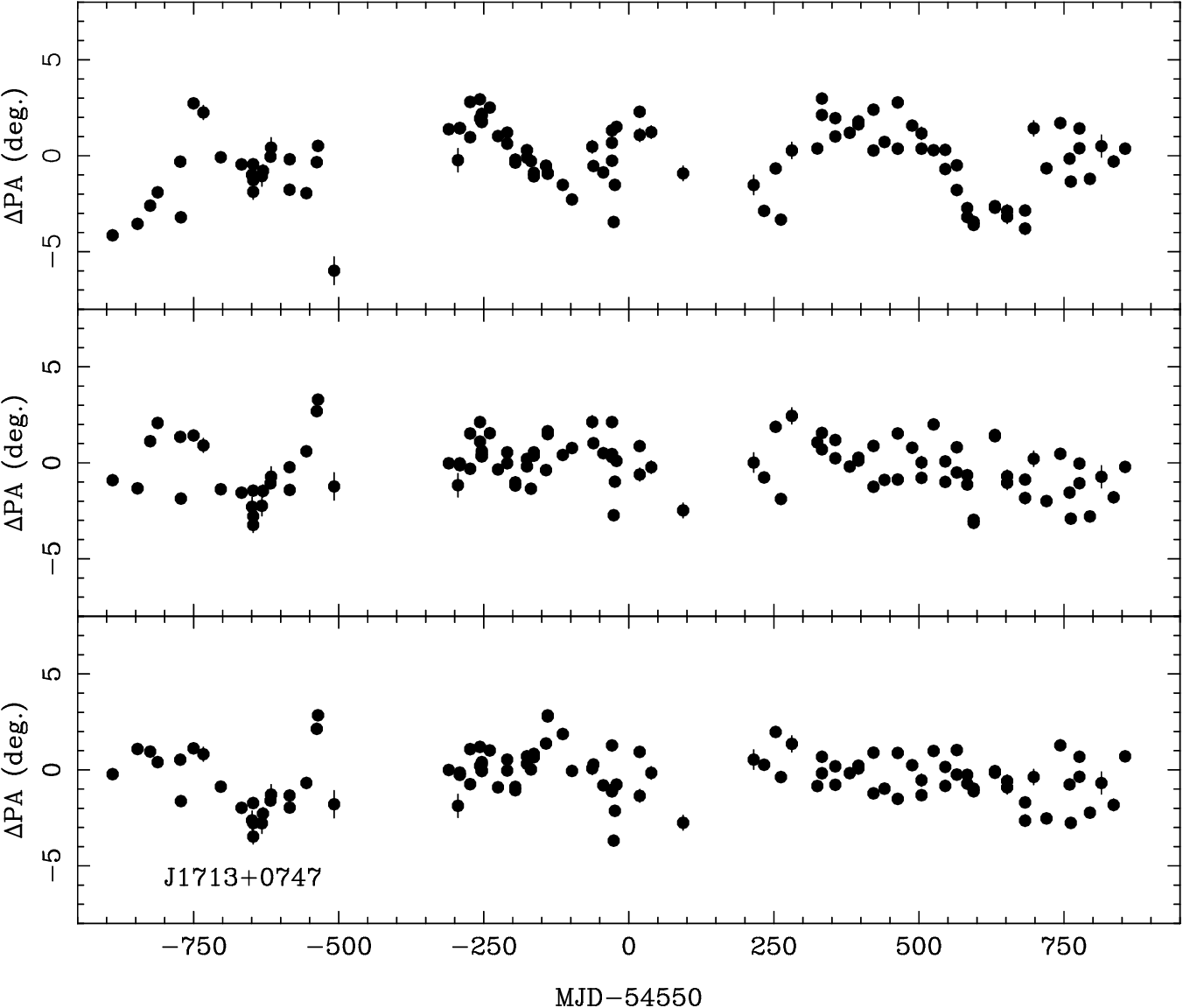}
\includegraphics[angle=-90,scale=.328]{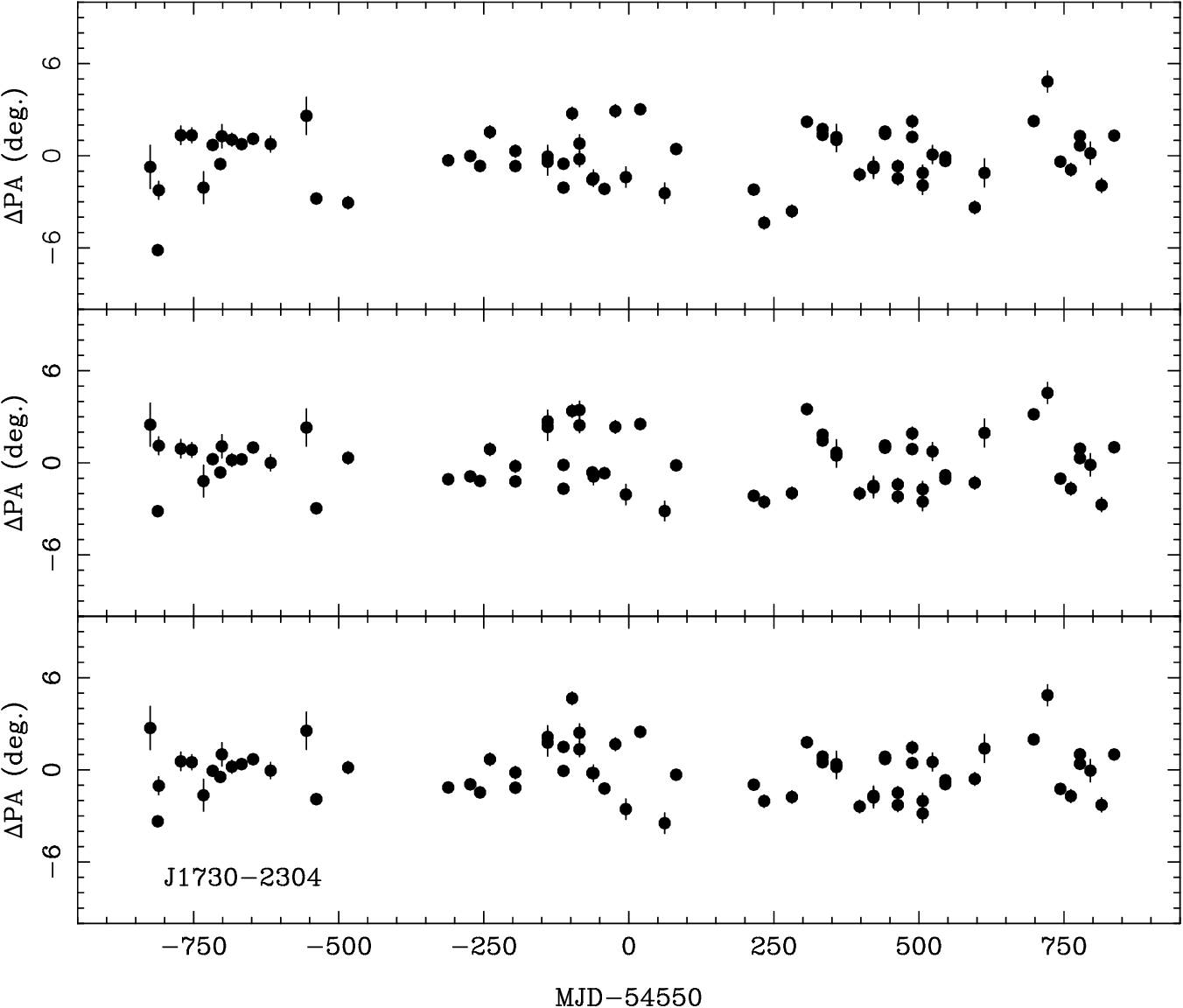}
\includegraphics[angle=-90,scale=.328]{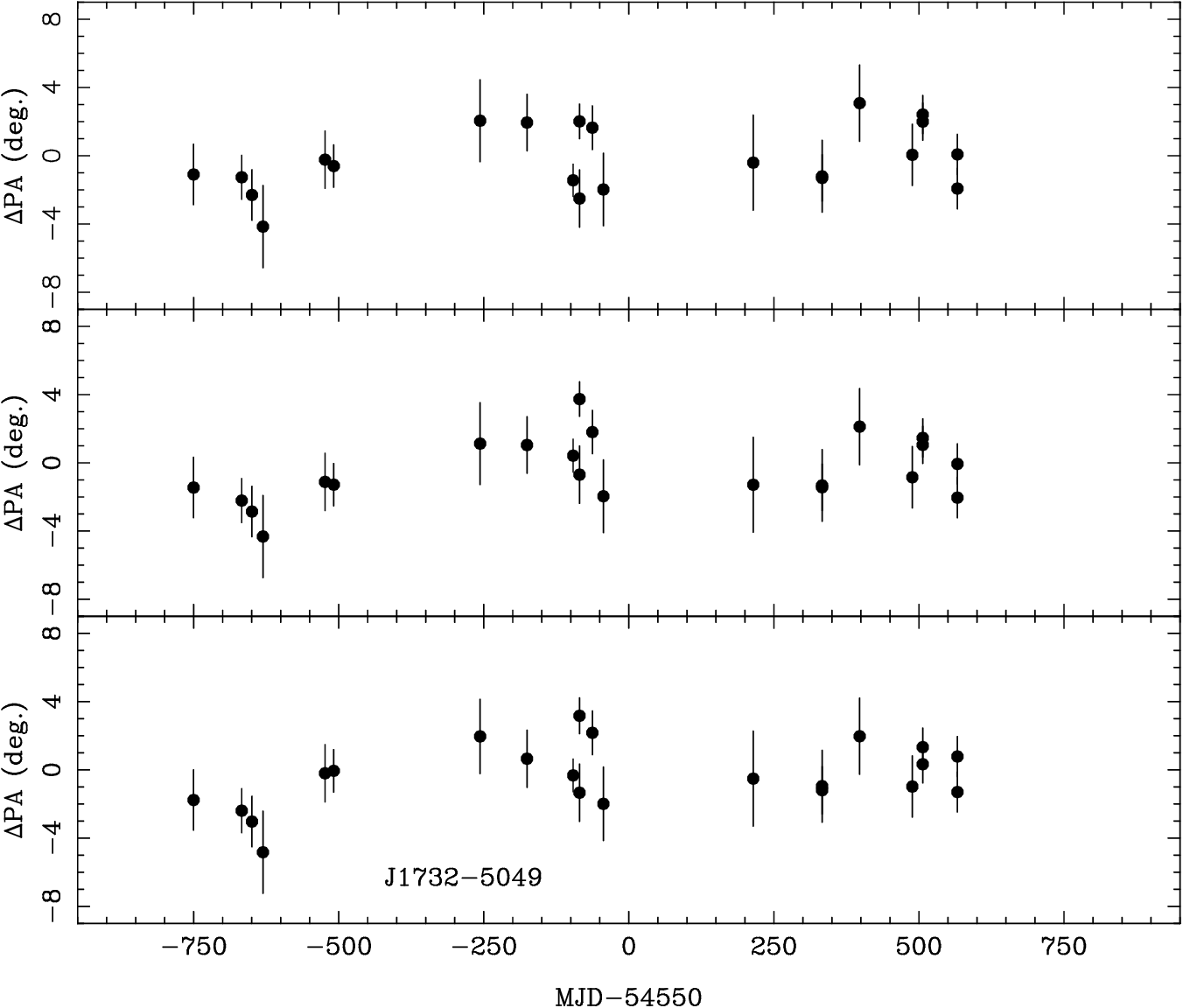}
\caption{PA differences as a function of time for six of the PPTA
  pulsars. See Figure~\ref{Fig:padiff1} for
  details.\label{Fig:padiff2}}
\end{figure*}

\begin{figure*}
\centering
\includegraphics[angle=-90,scale=.328]{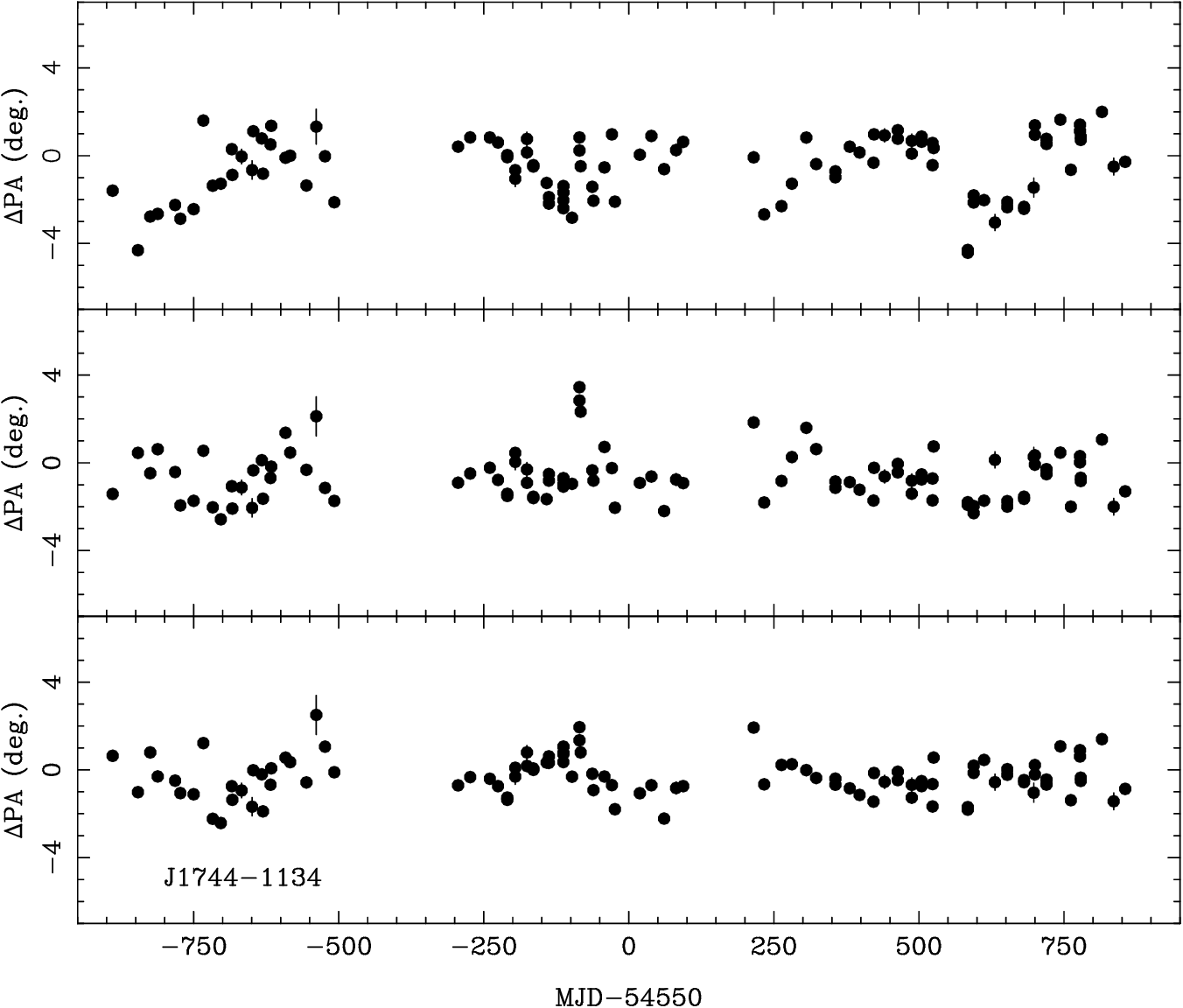}
\includegraphics[angle=-90,scale=.328]{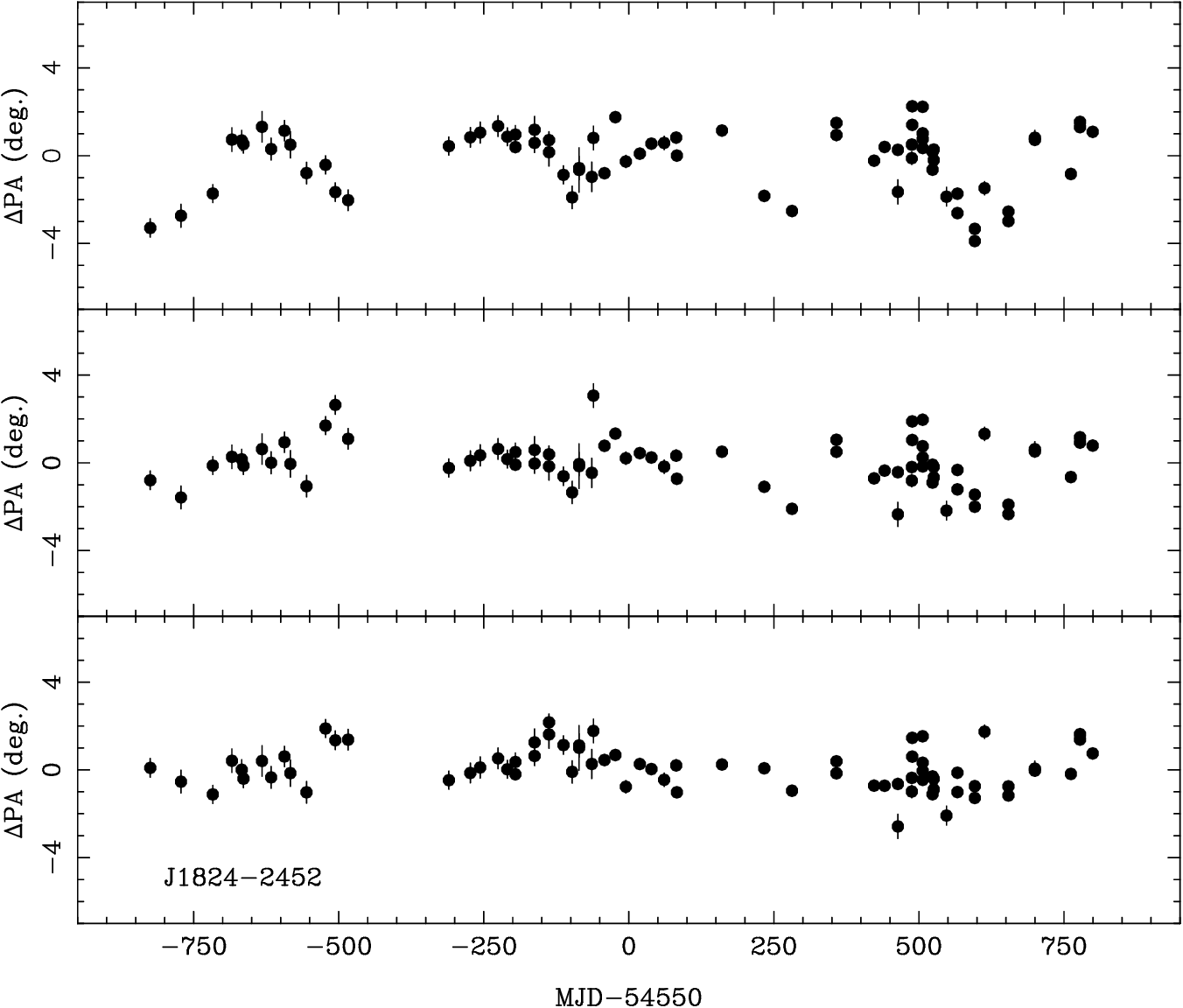}
\includegraphics[angle=-90,scale=.328]{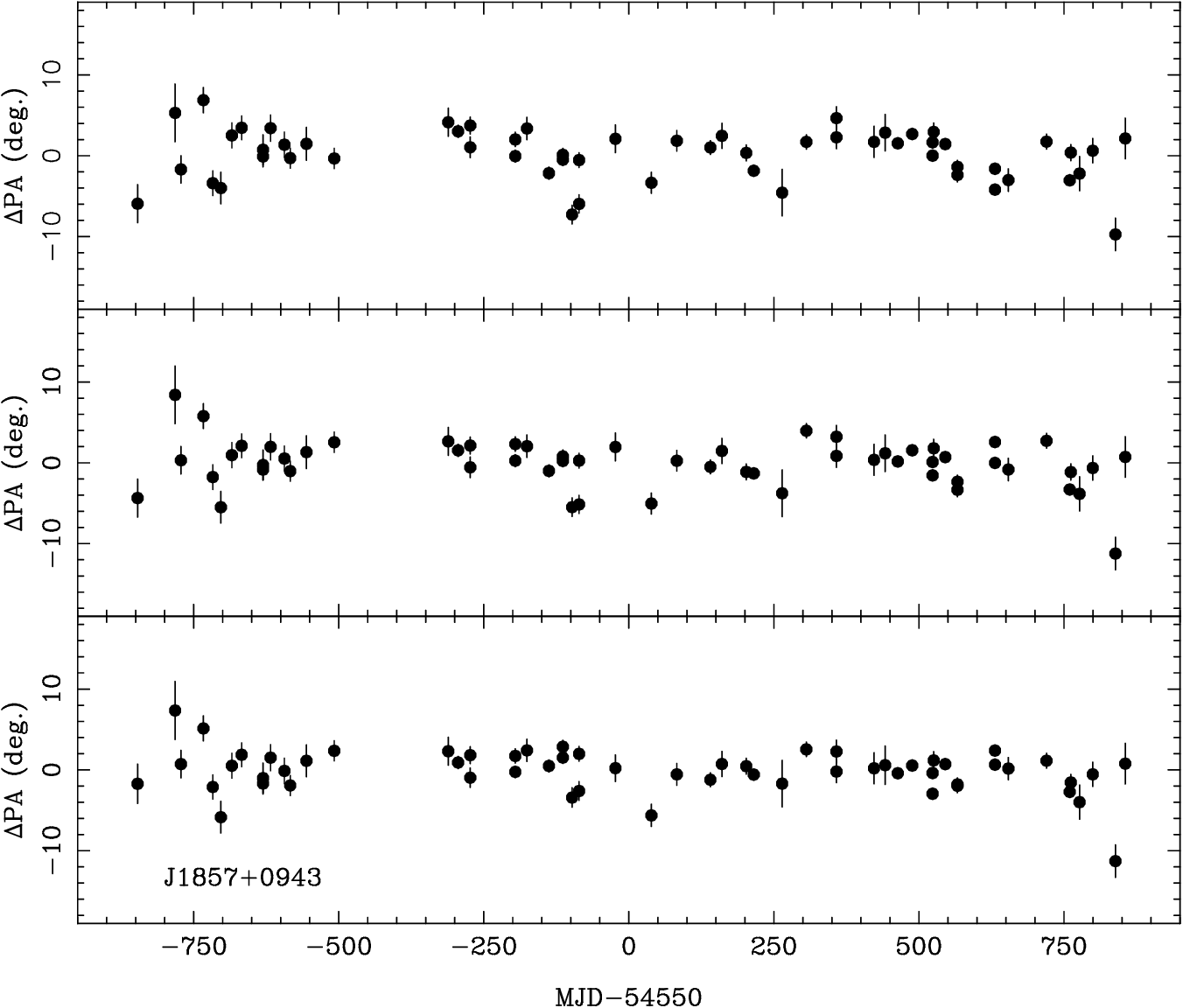}
\includegraphics[angle=-90,scale=.328]{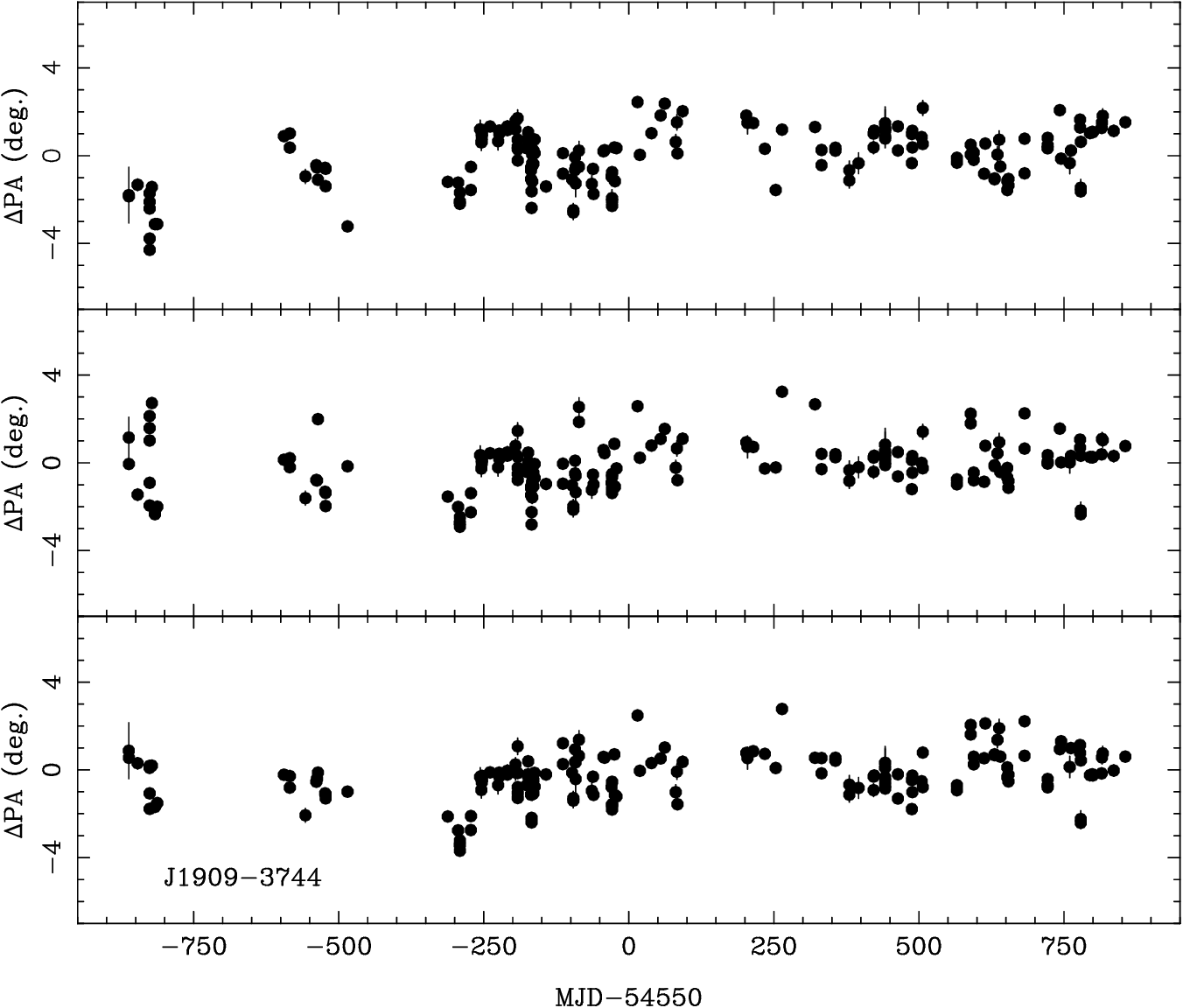}
\includegraphics[angle=-90,scale=.328]{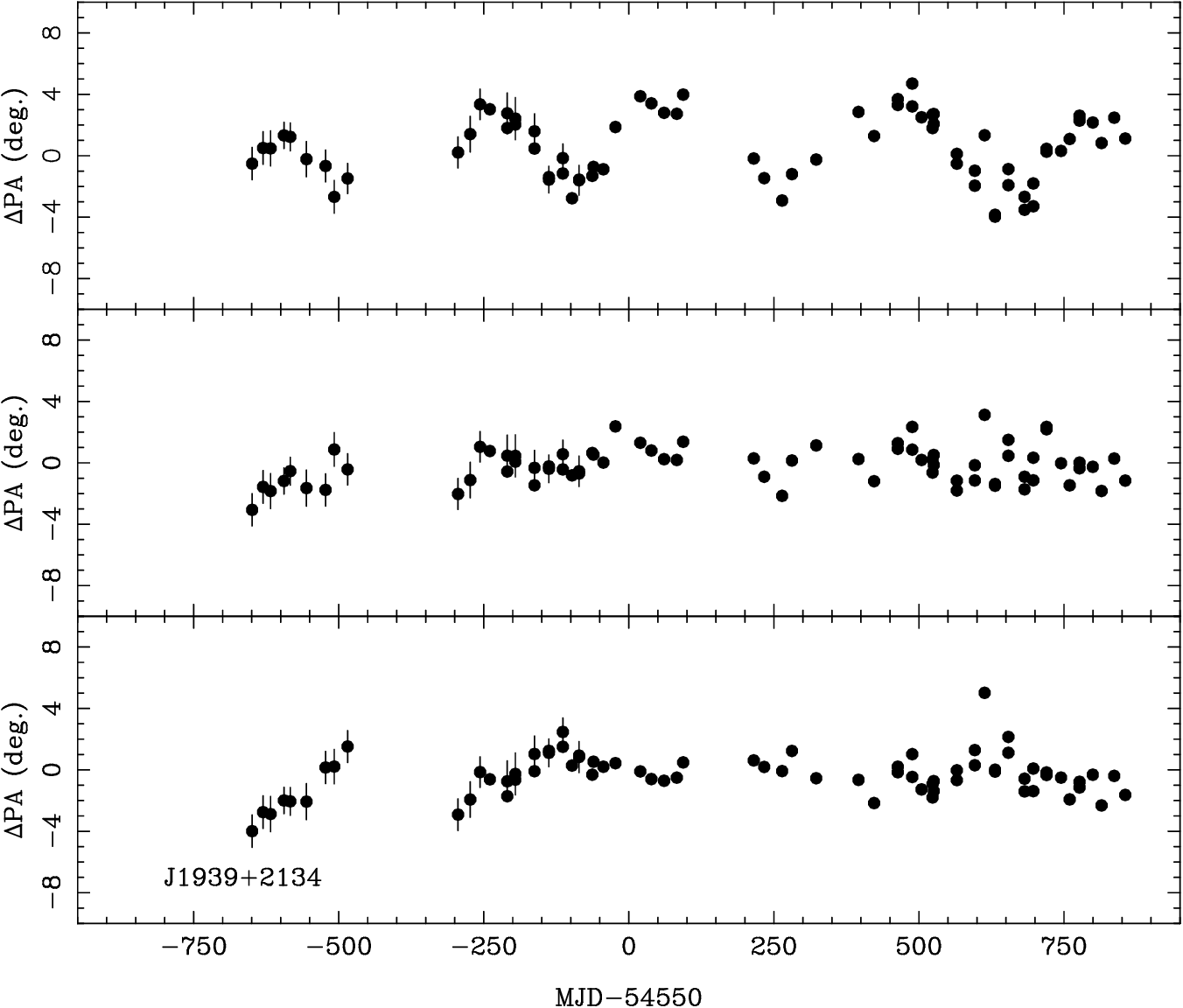}
\includegraphics[angle=-90,scale=.328]{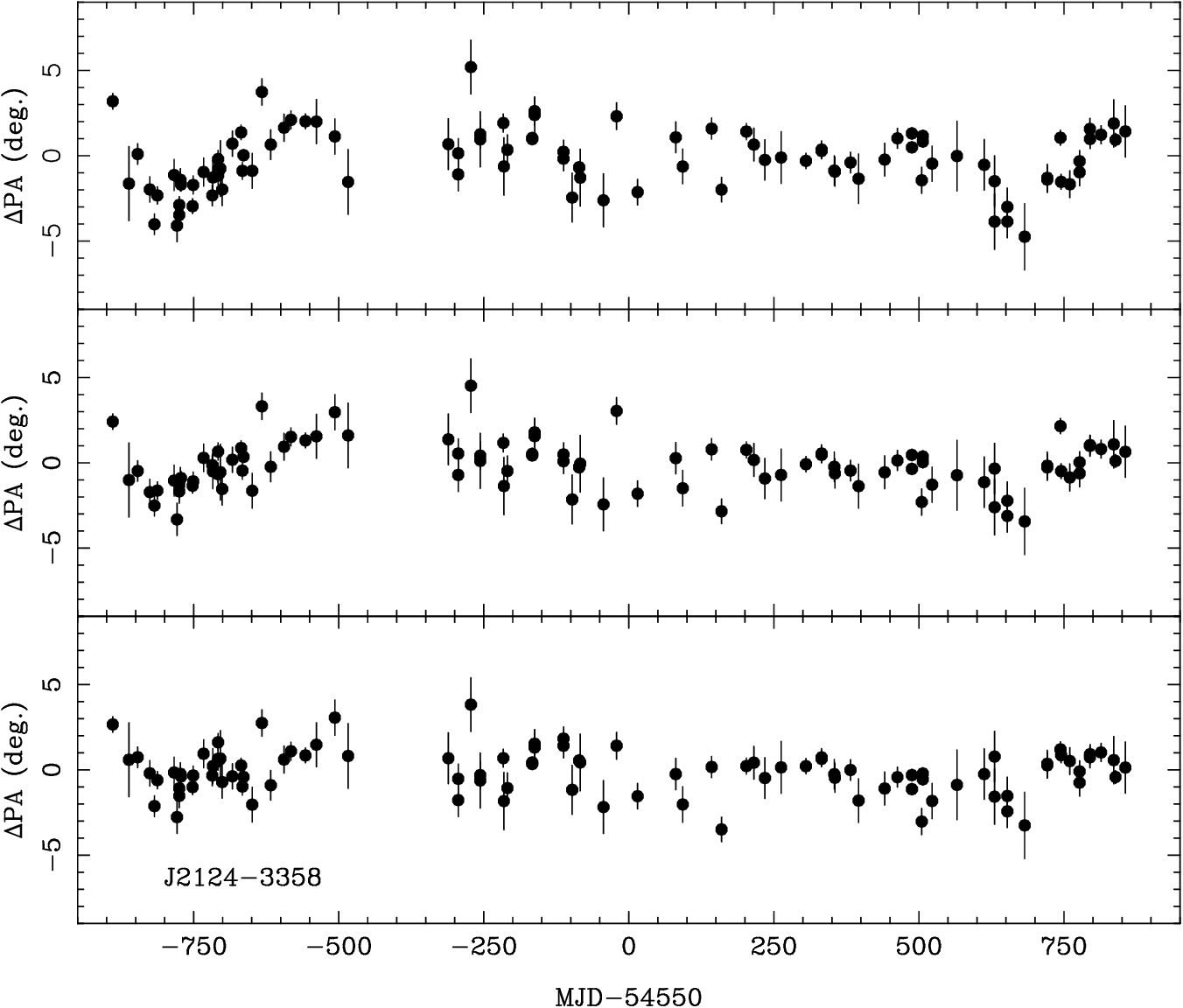}
\caption{PA differences as a function of time for six of the PPTA
  pulsars. See Figure~\ref{Fig:padiff1} for
  details.\label{Fig:padiff3}}
\end{figure*}

\begin{figure*}
\centering
\centering
\includegraphics[angle=-90,scale=.328]{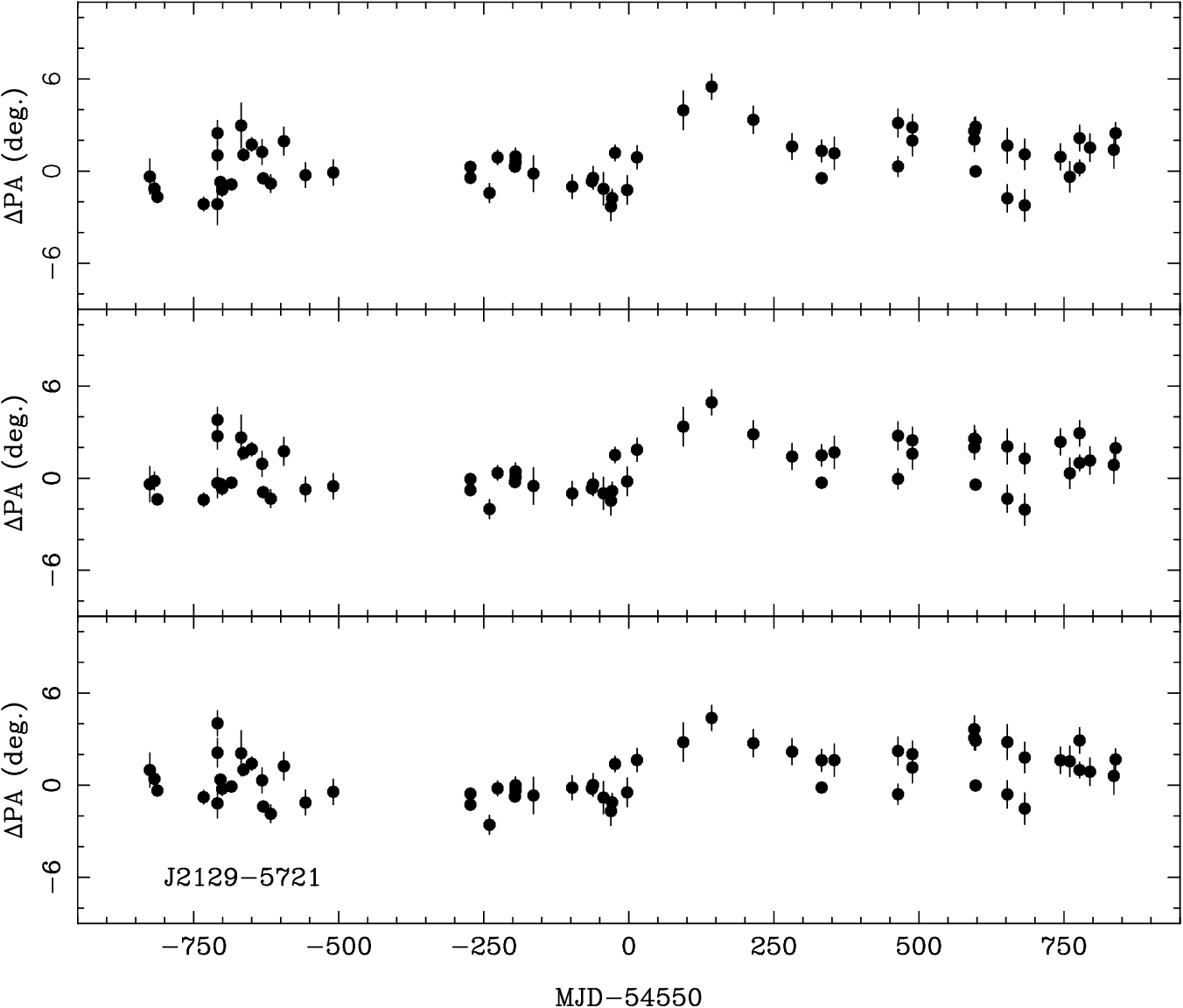}
\includegraphics[angle=-90,scale=.328]{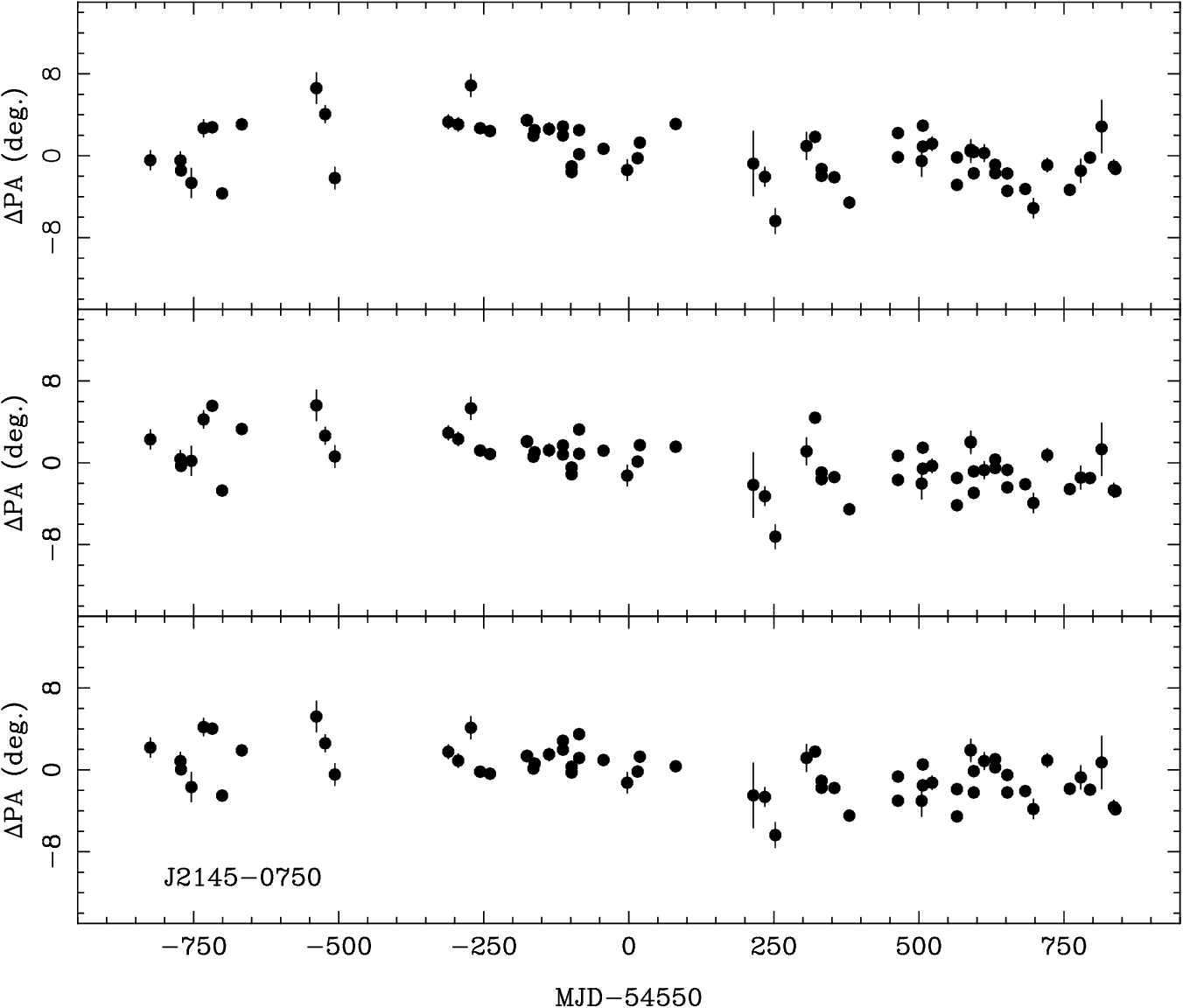}
\caption{PA differences as a function of time for two of the PPTA
  pulsars. See Figure~\ref{Fig:padiff1} for
  details.\label{Fig:padiff4}}
\end{figure*}

We have used these programs to compute RMi for each observation of
each pulsar. Observed PA differences have been corrected for the
rotation occurring in the ionosphere to give the PAs above
the ionosphere. In Figure \ref{Fig:padiff1} -- \ref{Fig:padiff4} we
show the observed PA differences in the upper panel for the 20 PPTA
pulsars. The middle panel gives the PA differences after correction
for ionospheric rotation using {\sc farrot} to compute RMi and the
lower panel shows the PA differences after correction using {\sc
  getRM\_iono}.

This figure shows that the PA variations are less after removal of the
ionospheric contribution. This is especially noticable for pulsars in
the northern sky, e.g., PSRs J1713+0747 and J1939+2134. For these
pulsars viewed from Parkes the line of sight to the pulsar is roughly
tangential to the local magnetic field (the dip angle at Parkes is
about $-60\degr$) and so the RMi is large. There is a clear annual
variation in the observed PAs resulting from the aliasing of the
diurnal ionospheric variation with the annual sidereal motion. This
annual variation is largely removed by the ionospheric correction.
Table~\ref{tb:pavar} gives the rms scatter in the PAs and the
corresponding reduced $\chi^2$ after a weighted fit of a straight line
to the corrected PAs for both the {\sc farrot} and IRI corrections for
the PPTA pulsars. Without exception the PAs after the IRI
corrections have a smaller rms scatter and reduced $\chi^2$ than those
after the {\sc farrot} correction.

\begin{table*}
\small
\centering
\begin{minipage}[]{120mm}

\caption[]{Rms variations and reduced $\chi^2$ for ionosphere-corrected PAs}
\label{tb:pavar}\end{minipage}
\tabcolsep 3mm
\begin{tabular}{ccccc}
\hline\noalign{\smallskip}
PSR  &\multicolumn{2}{c}{\sc farrot} & \multicolumn{2}{c}{\sc IRI}\\
  &  $\sigma_{PA}$ & Reduced &  $\sigma_{PA}$ & Reduced \\
  & ($\degr$)   & $\chi^2$  & ($\degr$) & $\chi^2$   \\
\hline\noalign{\smallskip}
J0437$-$4715  &1.14   &833.3   &0.96    &488.6  \\
J0613$-$0200  &2.31   &5.9     &2.15    &4.7    \\
J0711$-$6830  &1.82   &3.5     &1.70    &3.3  \\
J1022$+$1001    &1.22   &192.8   &0.93    &96.5 \\
J1024$-$0719  &1.41   &47.3    &1.31    &43.1 \\\\
J1045$-$4509  &1.74   &5.6     &1.63    &4.6 \\
J1600$-$3053  &1.33   &10.6    &1.08    &6.5 \\
J1603$-$7202  &1.35   &4.0     &1.29    &2.9 \\
J1643$-$1224  &1.43   &5.7     &1.09    &3.5 \\
J1713$+$0747    &1.40   &76.9    &1.37    &70.7 \\\\
J1730$-$2304  &1.83   &25.2    &1.71    &17.0 \\
J1732$-$5049  &1.93   &1.8     &1.92    &1.5 \\
J1744$-$1134  &1.28   &779.9   &0.98    &411.9 \\
J1824$-$2452  &1.08   &16.5    &0.95    &11.1 \\
J1857$+$0943    &3.02   &5.3     &2.71    &4.0 \\\\
J1909$-$3744  &1.15   &138.1   &1.11    &123.5 \\
J1939$+$2134    &1.22   &64.3    &1.42    &50.6 \\
J2124$-$3358  &1.44   &3.4     &1.31    &3.0 \\
J2129$-$5721  &1.72   &5.4     &1.71    &5.2 \\
J2145$-$0750  &2.49   &31.8    &2.31    &20.2 \\
\noalign{\smallskip}\hline
\end{tabular}
\end{table*}

\section{Results and Discussion}
\label{sect:disc}
After the ionospheric correction, the PA variations for essentially
all of the PPTA pulsars are dominated by a short-term scatter.  As
shown by the large values of reduced $\chi^2$ (Table~\ref{tb:pavar}),
this scatter is much larger than the nominal uncertainty of the PAs in
most cases. For the stronger pulsars, the PA uncertainty estimated
from the off-pulse rms variations in the Stokes profiles is very
small, often $\ll 1\degr$.  Residual calibration errors can exceed the
rms noise in the stronger pulsars, leading to PA offsets and hence
scatter in the PAs. Low-level un-excised RFI can affect both the
accuracy of the calibration and the Stokes data themselves also
leading to PA scatter. Finally, although the ionospheric correction
has reduced the short-term scatter in the PAs, it cannot be completely
accurate as it relies on models for the ionosphere and its variations
whereas the actual ionosphere is dynamic and changes on short
timescales in often unpredictable ways. It is difficult to separate
these contributions to the scatter in the corrected PAs, but it is
probable that this last effect is dominant. Near real-time monitoring
of the total electron content, for example, using the dual-frequency
Global Positioning System \citep[e.g.,][]{ggt+08}, should help to
improve the accuracy of the RMi measurements and hence reduce the
overall short-term scatter in the PAs.

Figures \ref{Fig:padiff1} -- \ref{Fig:padiff4} show that there is little
if any significant long-term variation in the corrected PAs. This
implies that the mean PA of the emitted pulses and hence the
orientation of the spin and magnetic axes of the pulsar have varied by
$\lapp 1\degr$ over the nearly 5-year data span. This is not
surprising since six of the 20 pulsars are isolated and none of the 14
binary pulsars is in a close orbit where geodetic precession might be
expected \citep{bkk+08,mks+10}. In addition, most of the binary
companions are evidently of relatively low mass ($\lapp
0.5$~M$_\odot$). Because of the great age of these MSPs, any free
precession resulting from interactions or deformations in the spin-up
phase would be likely to have dissipated by now \citep[cf.][]{gaj09}.

\begin{figure*}
\centering
\includegraphics[angle=-90,scale=.328]{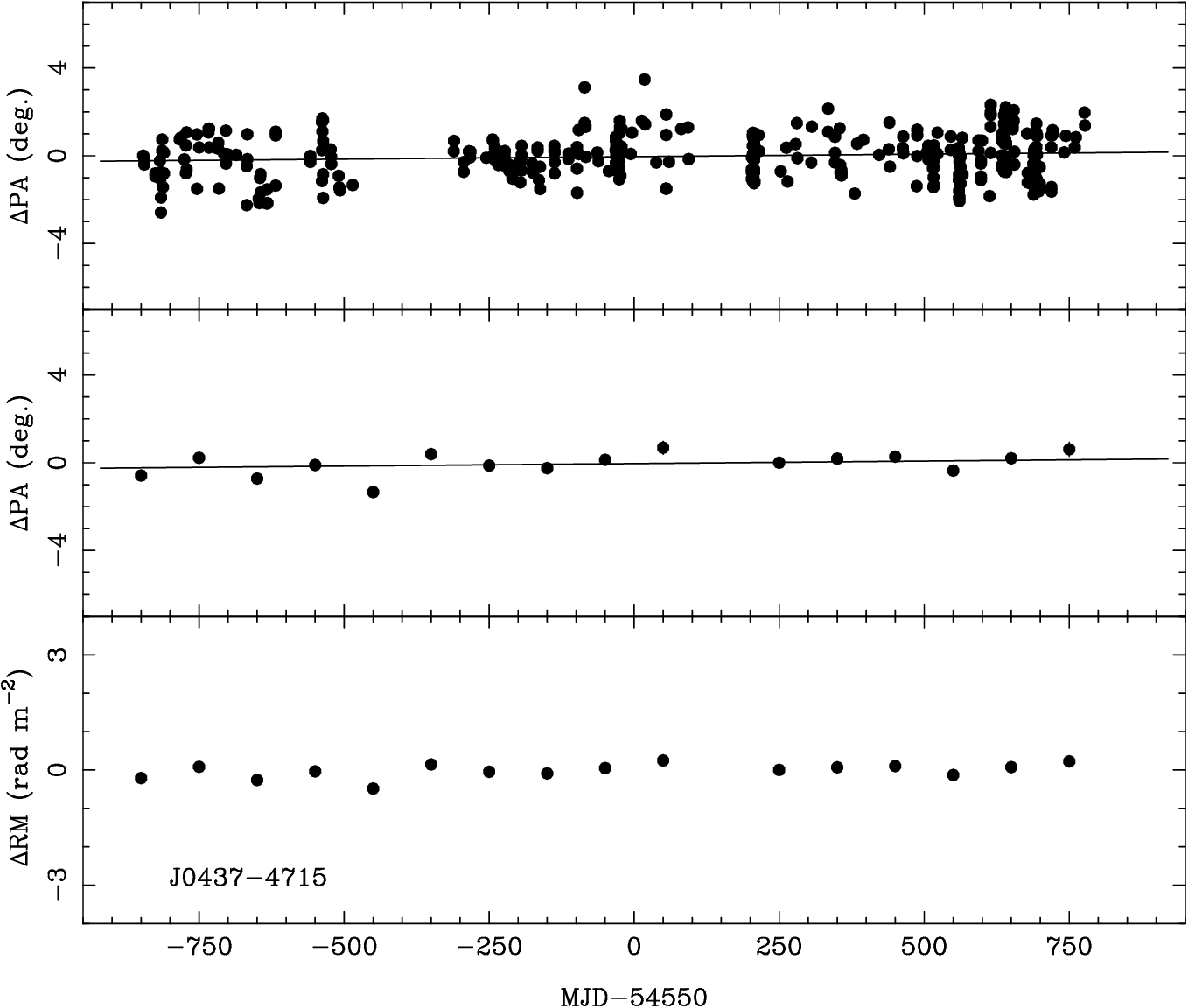}
\includegraphics[angle=-90,scale=.328]{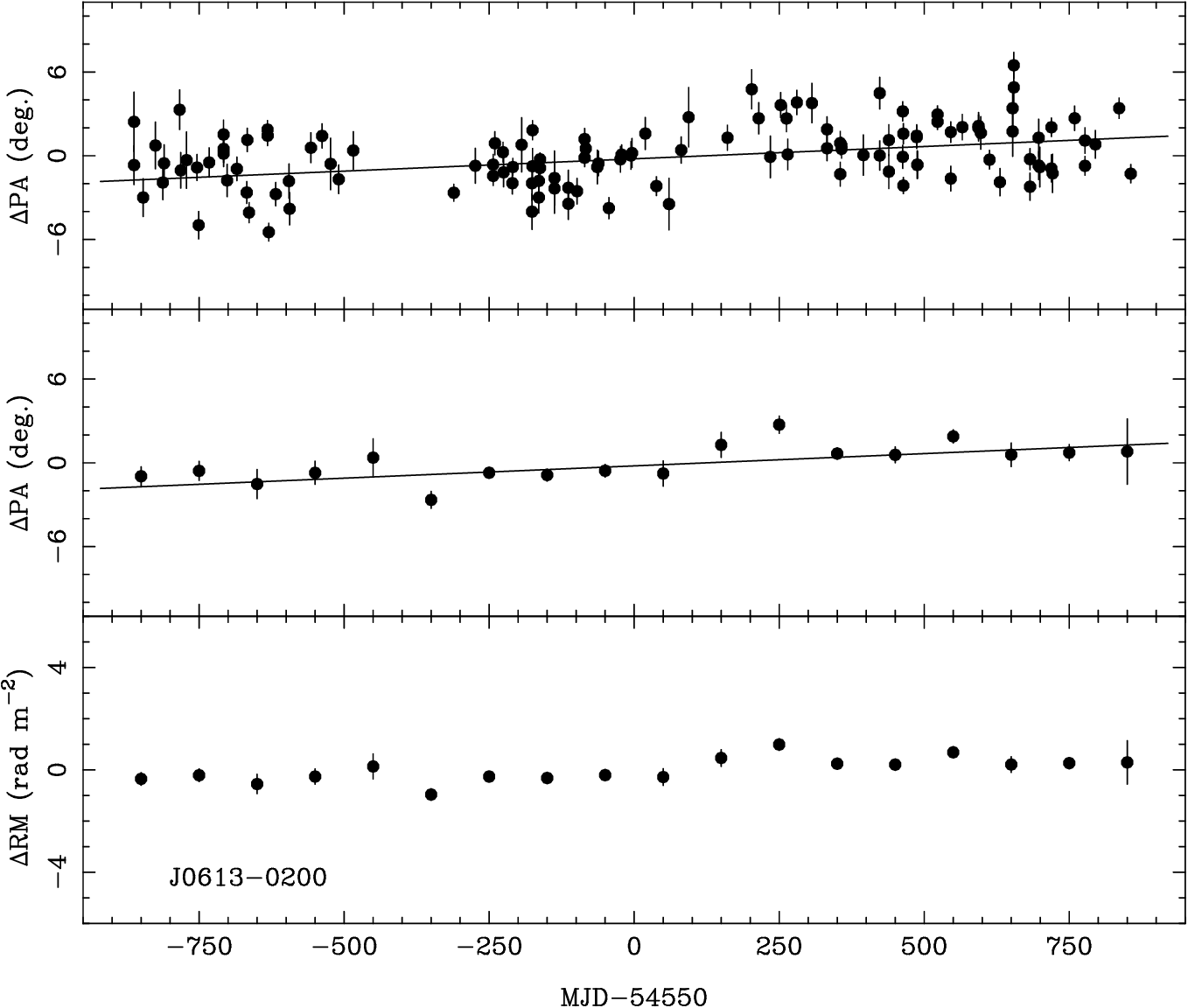}
\includegraphics[angle=-90,scale=.328]{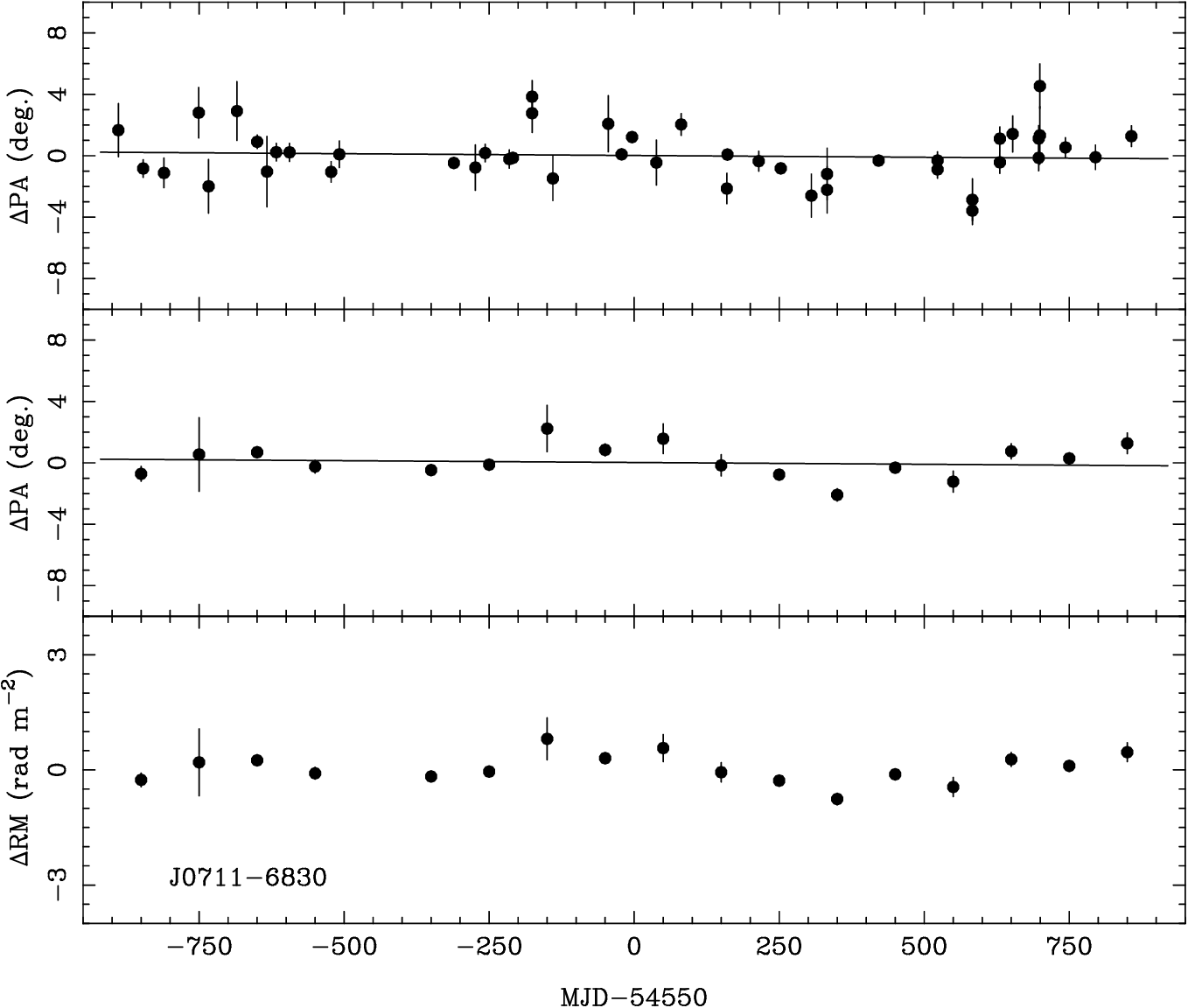}
\includegraphics[angle=-90,scale=.328]{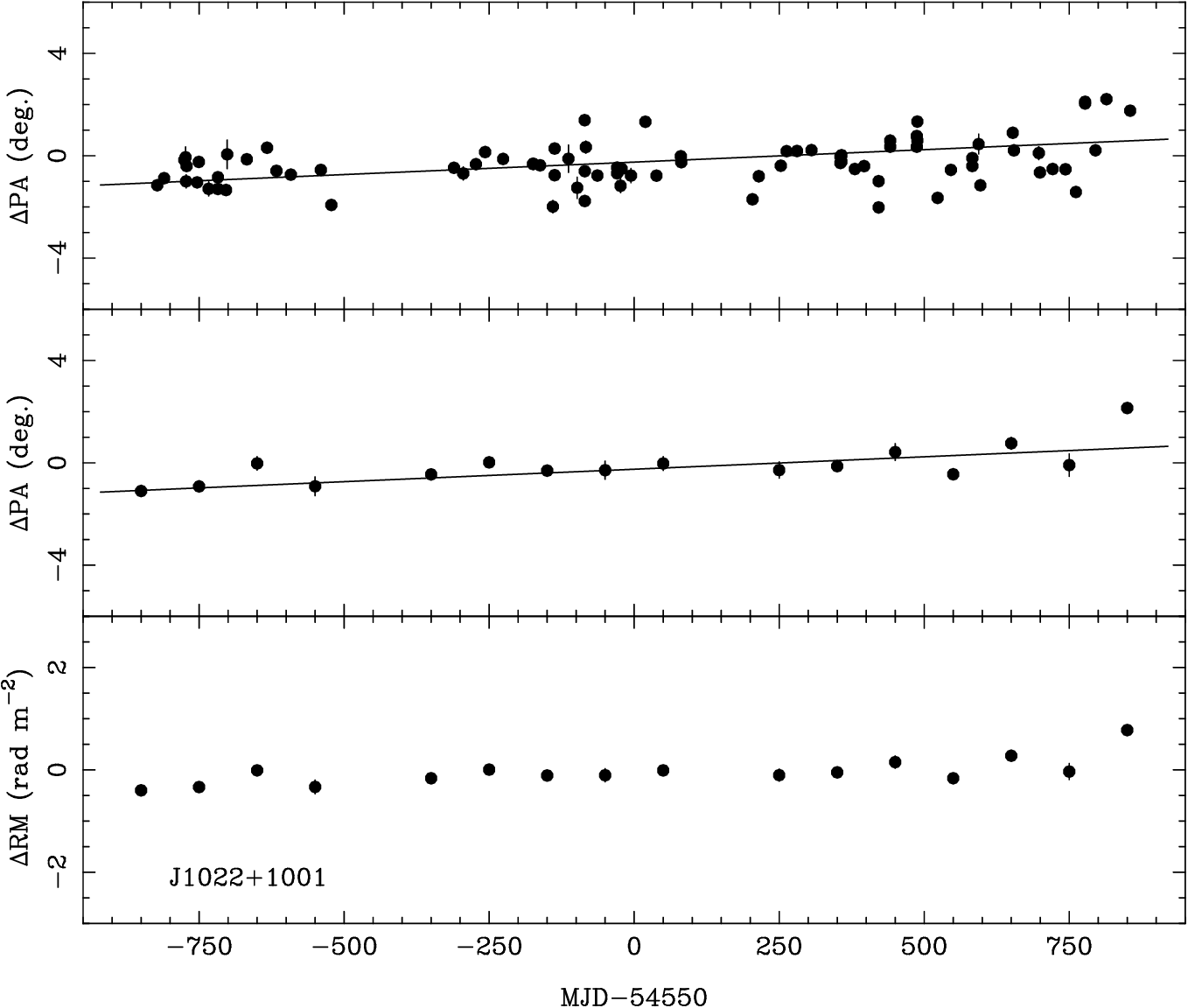}
\includegraphics[angle=-90,scale=.328]{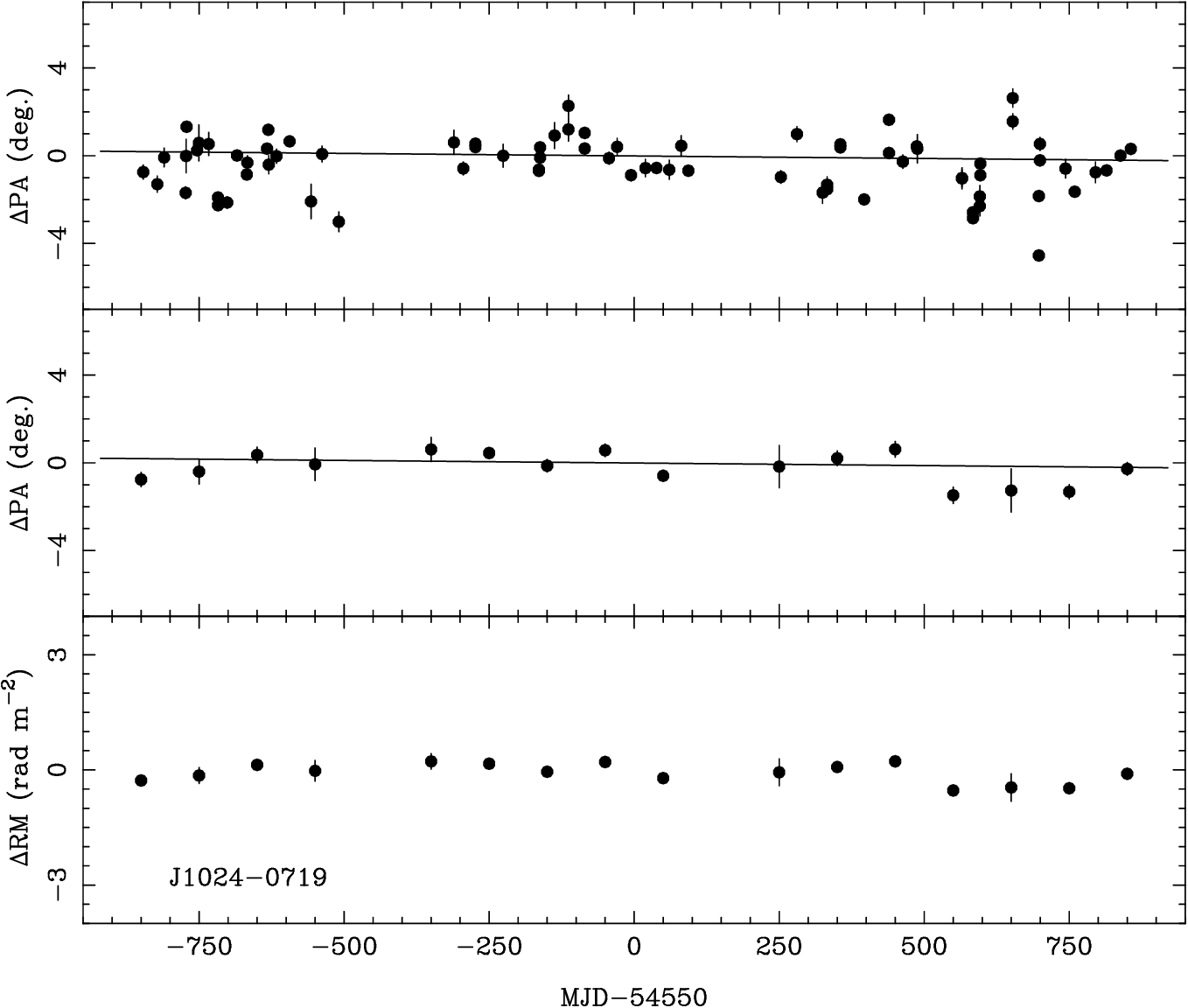}
\includegraphics[angle=-90,scale=.328]{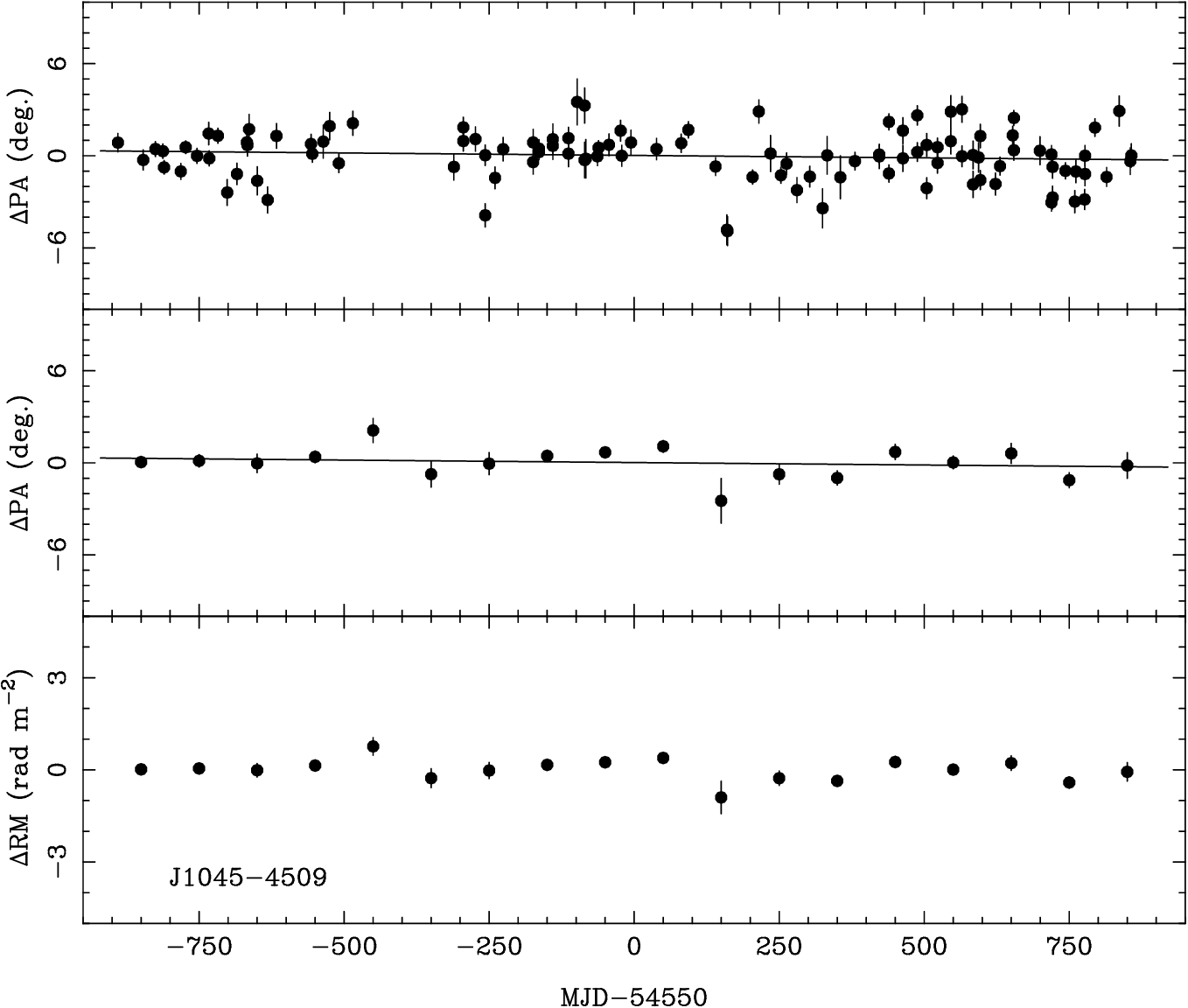}
\caption{PA variations for six of the PPTA pulsars. The upper
  panel of each sub-figure has the IRI-corrected PA variations showing
  the fitted linear trend. The middle panel shows the same data averaged
  over 100-day blocks with the fitted line as in the upper panel. The lower panel shows the
  smoothed variations converted to RM changes using \eqref{eq:fara}.
  \label{Fig:pavar1}}
\end{figure*}

\begin{figure*}
\centering
\includegraphics[angle=-90,scale=.328]{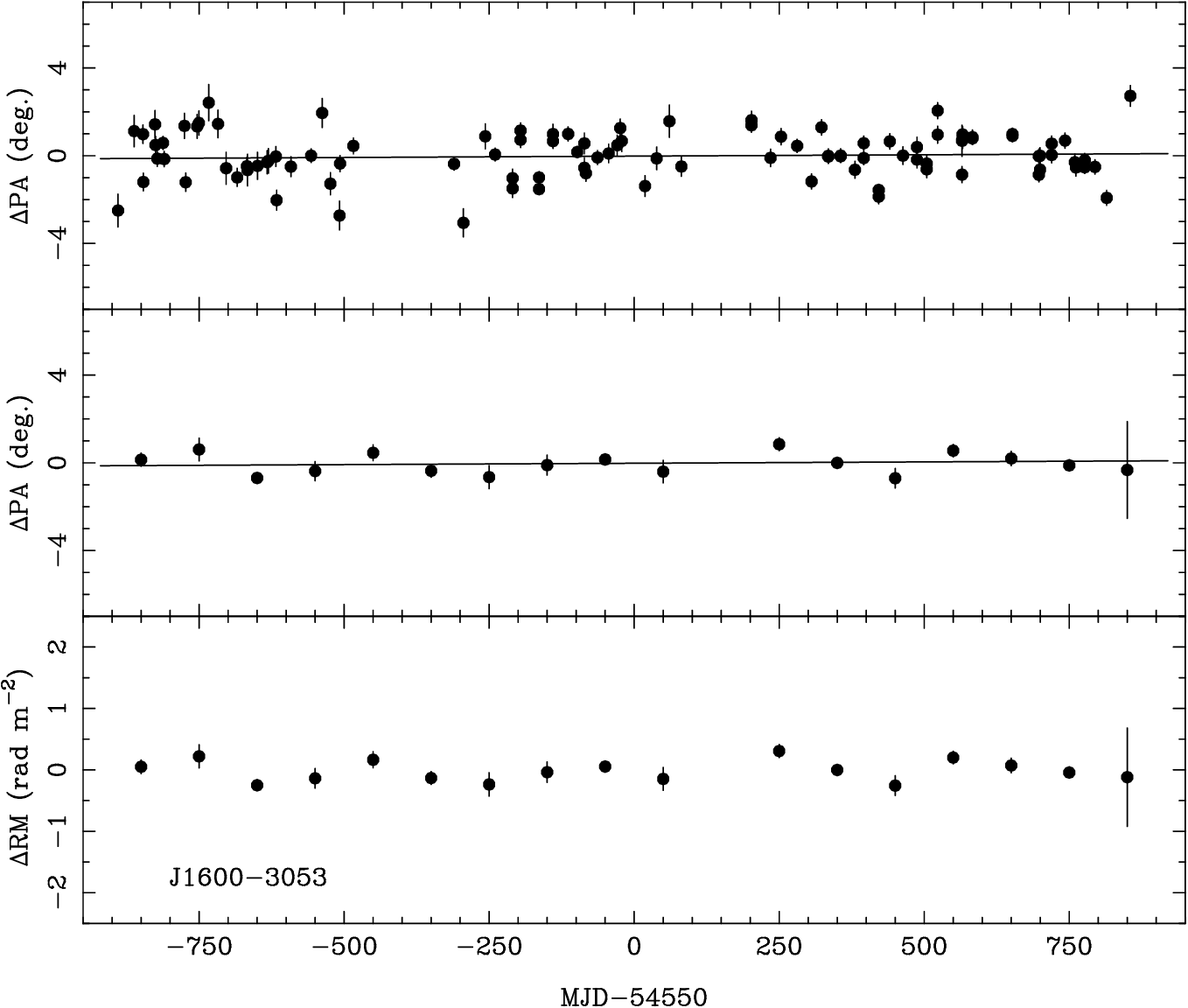}
\includegraphics[angle=-90,scale=.328]{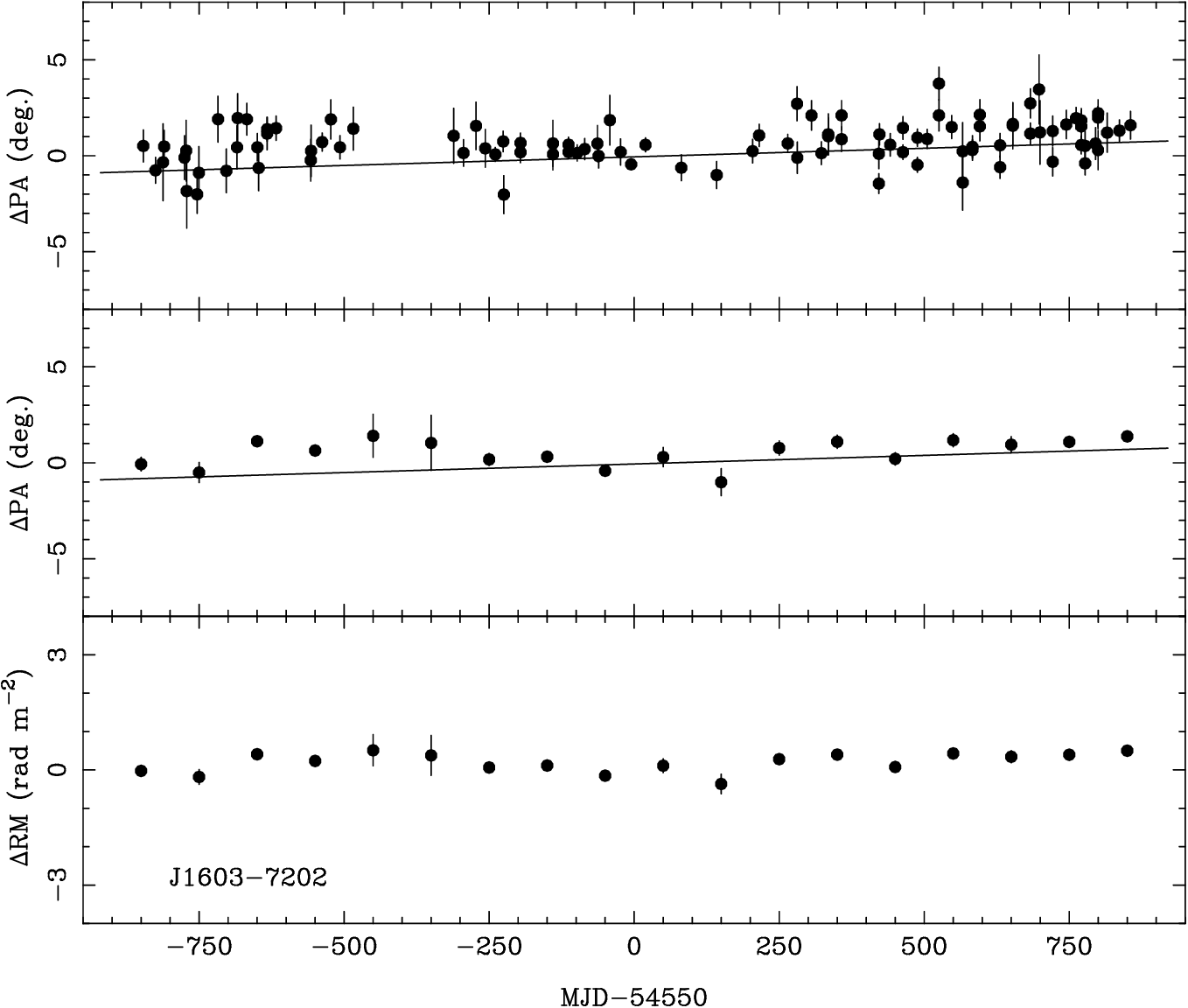}
\includegraphics[angle=-90,scale=.328]{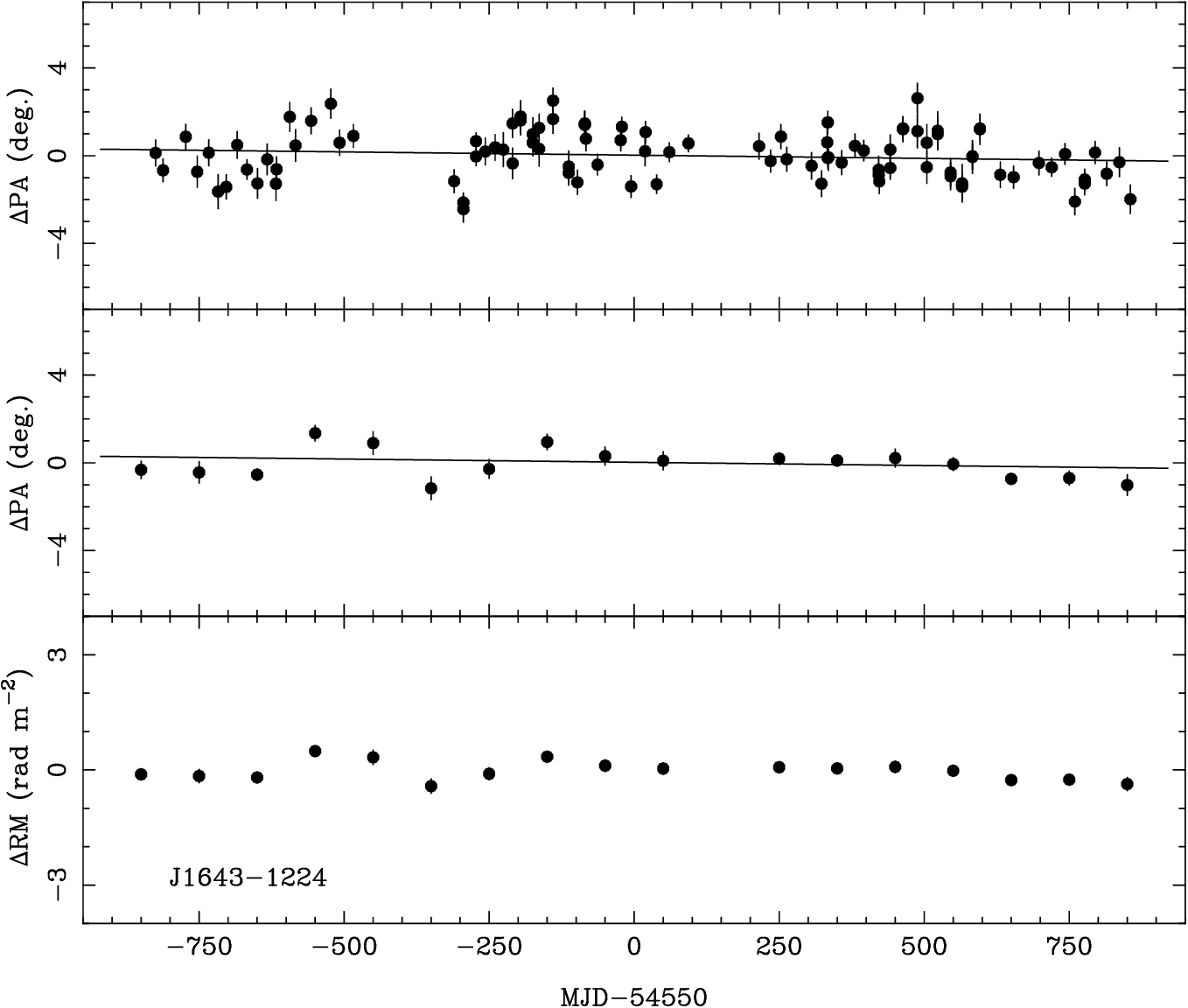}
\includegraphics[angle=-90,scale=.328]{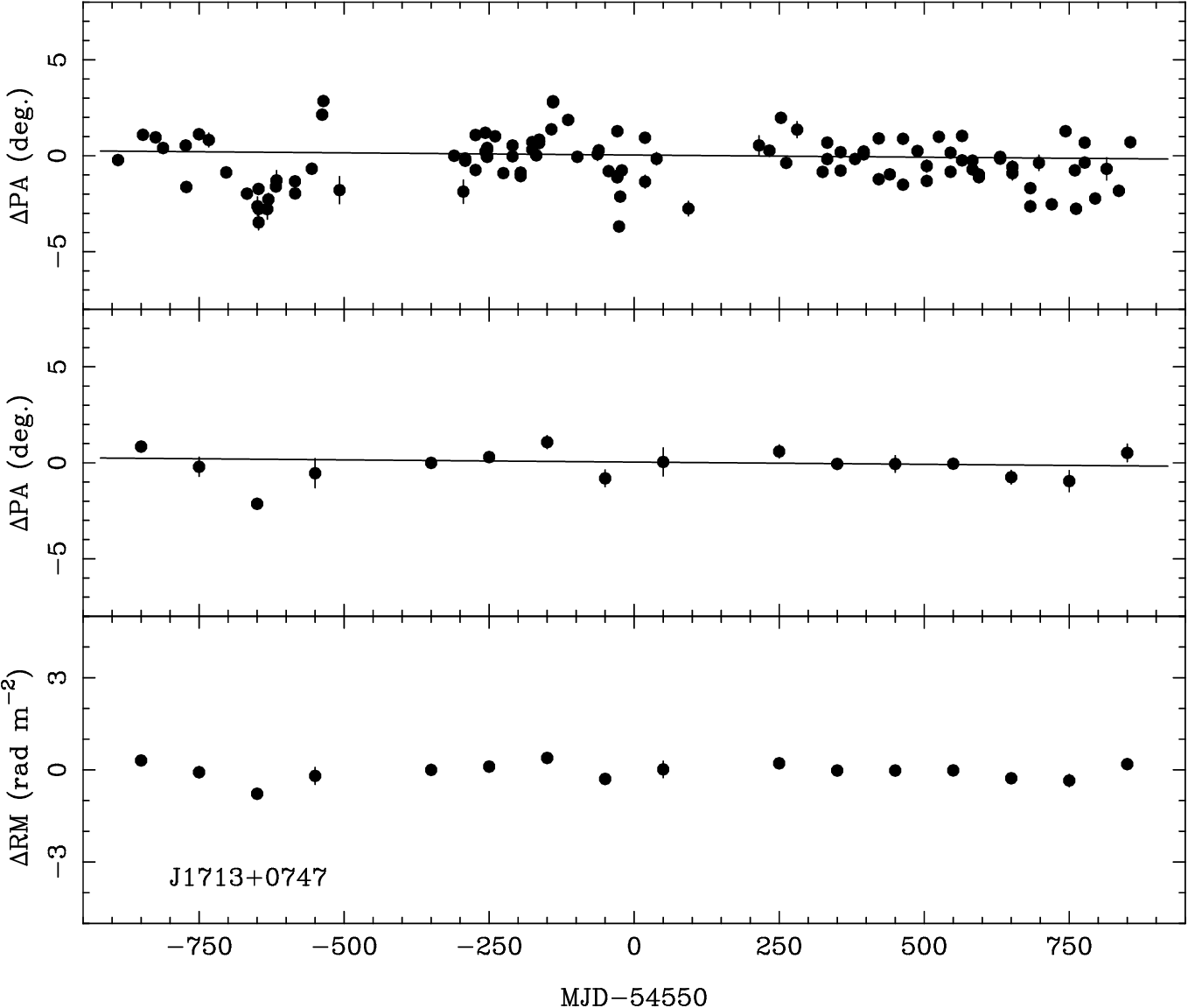}
\includegraphics[angle=-90,scale=.328]{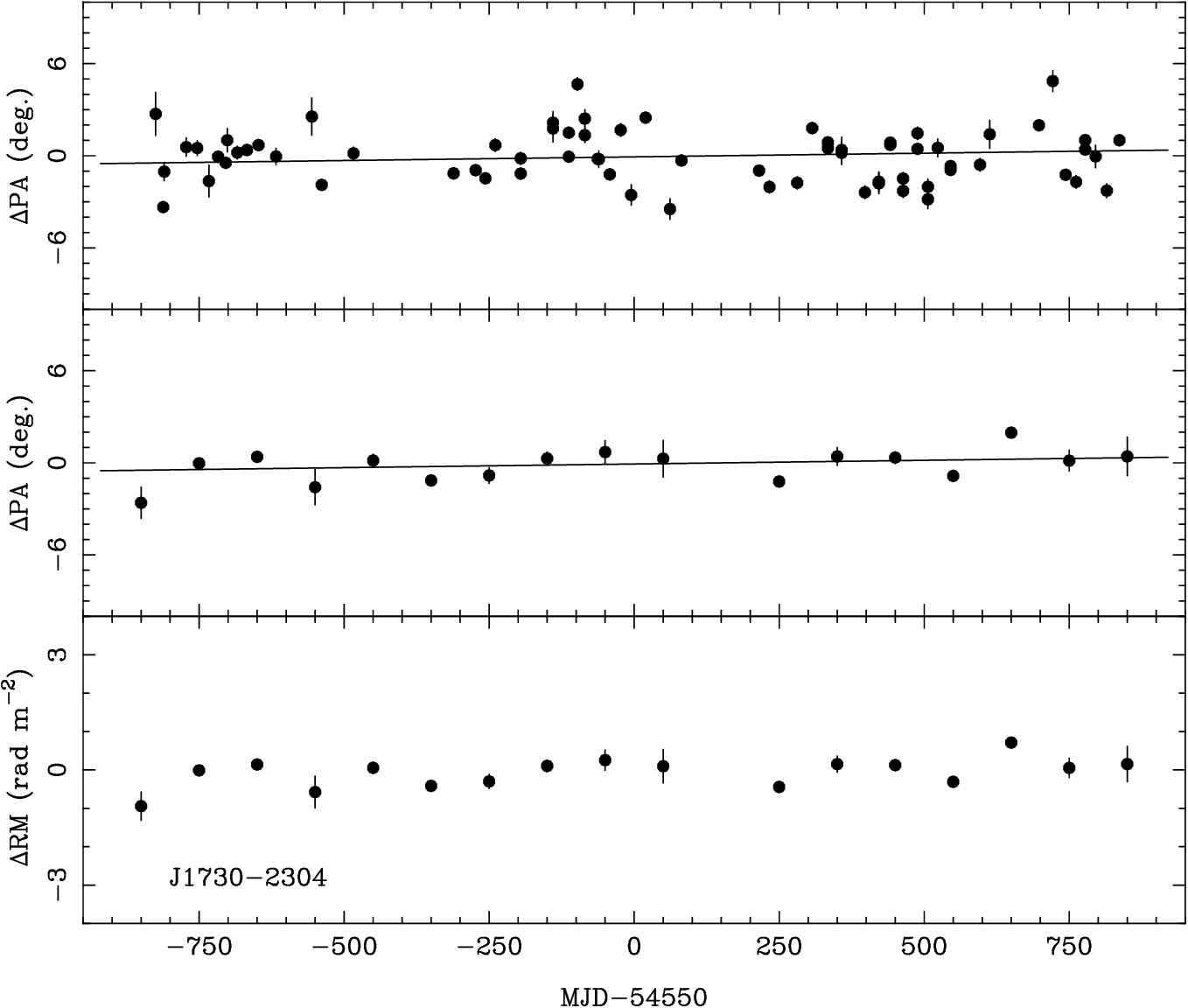}
\includegraphics[angle=-90,scale=.328]{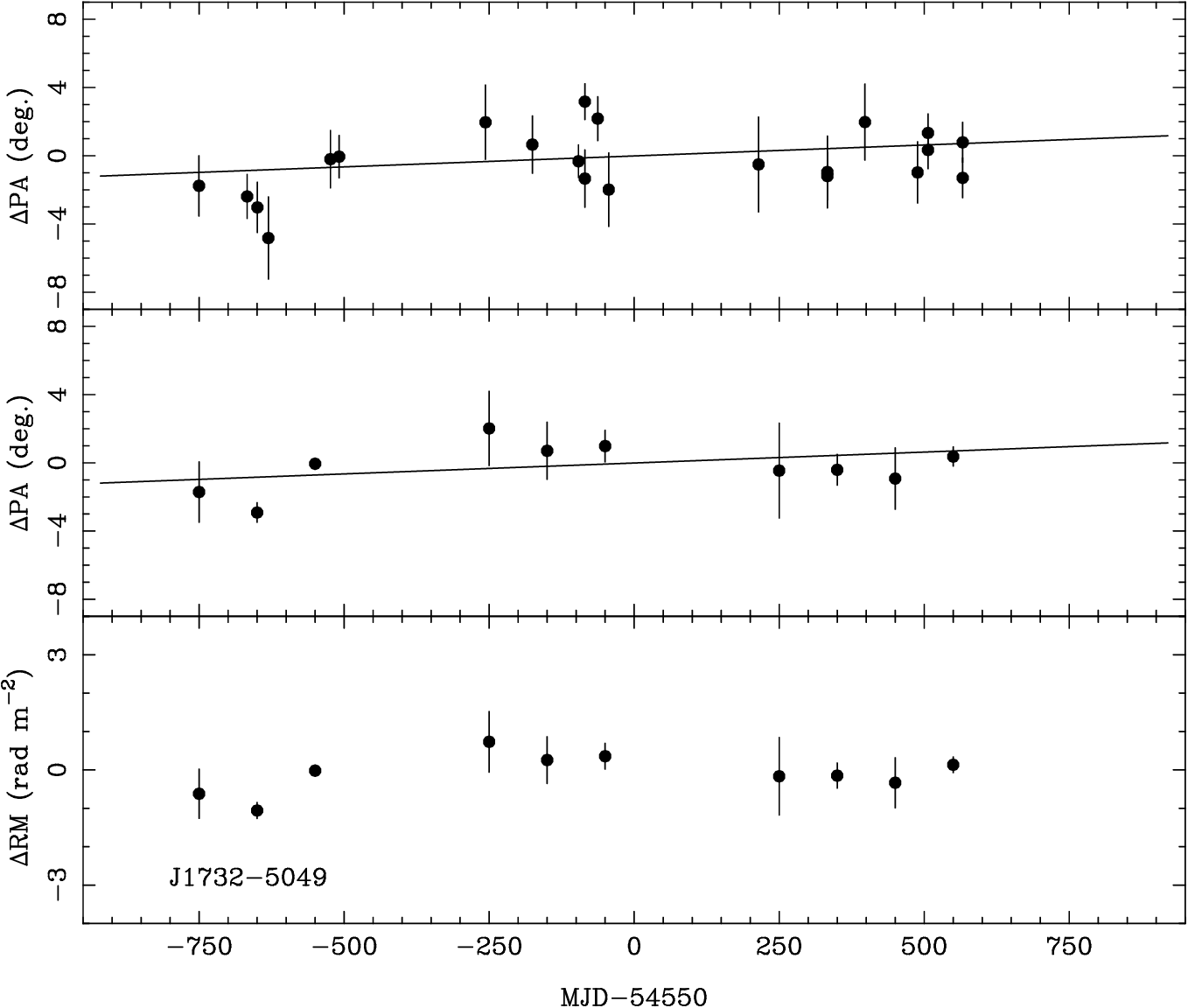}
\caption{PA variations for six of the PPTA pulsars. 
See Figure~\ref{Fig:pavar1} for
  details.\label{Fig:pavar2}}
\end{figure*}

\begin{figure*}
\centering
\includegraphics[angle=-90,scale=.328]{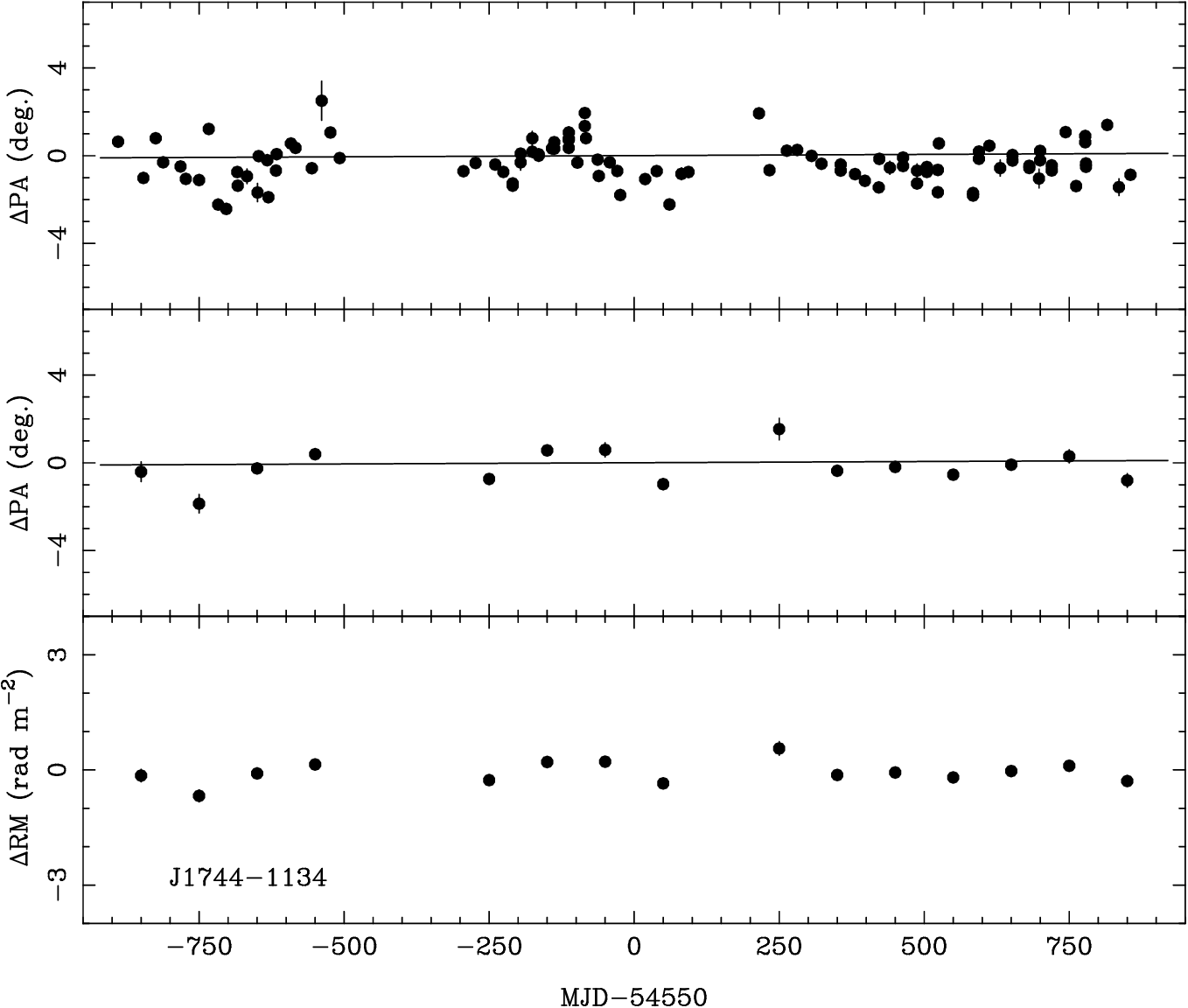}
\includegraphics[angle=-90,scale=.328]{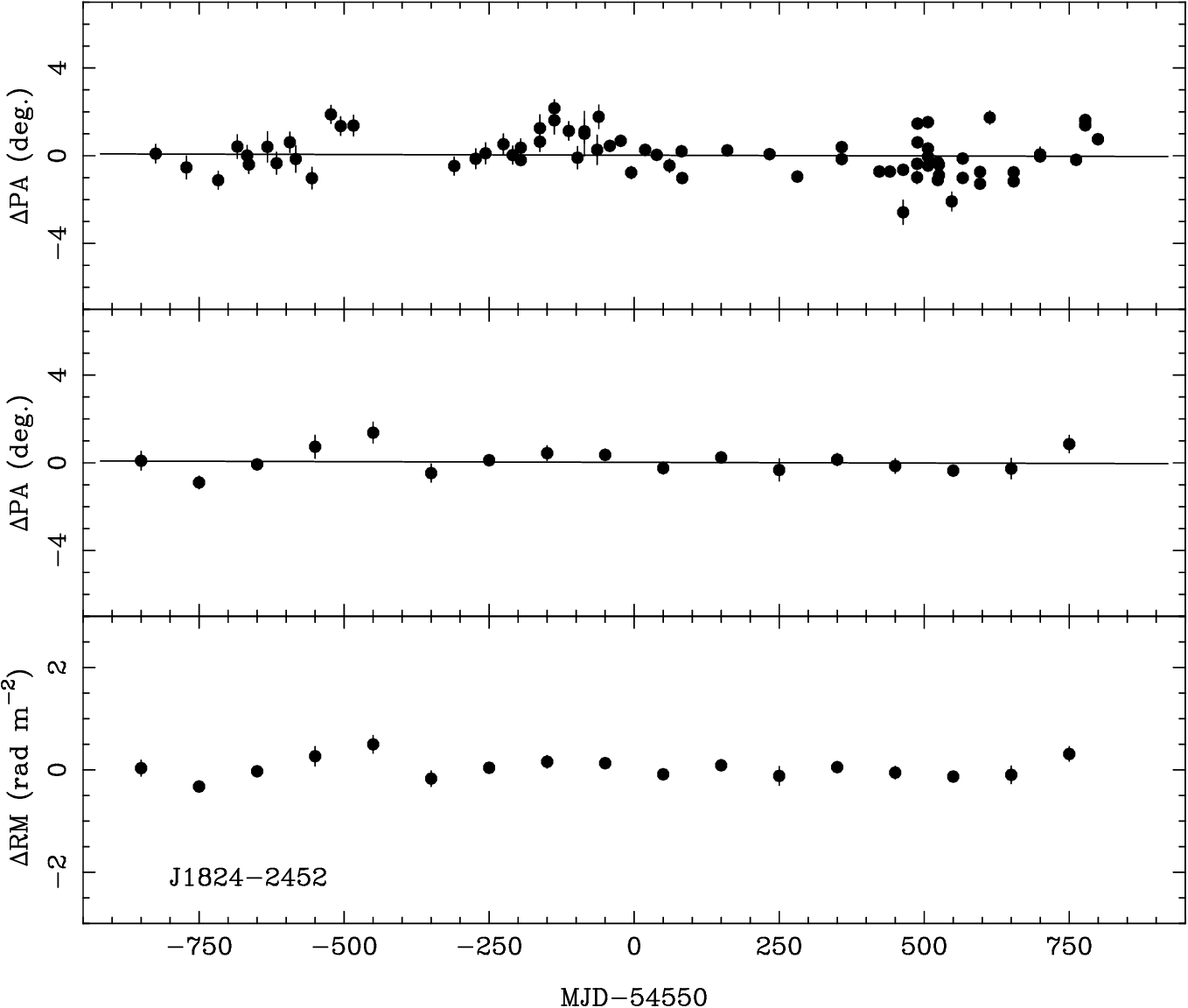}
\includegraphics[angle=-90,scale=.328]{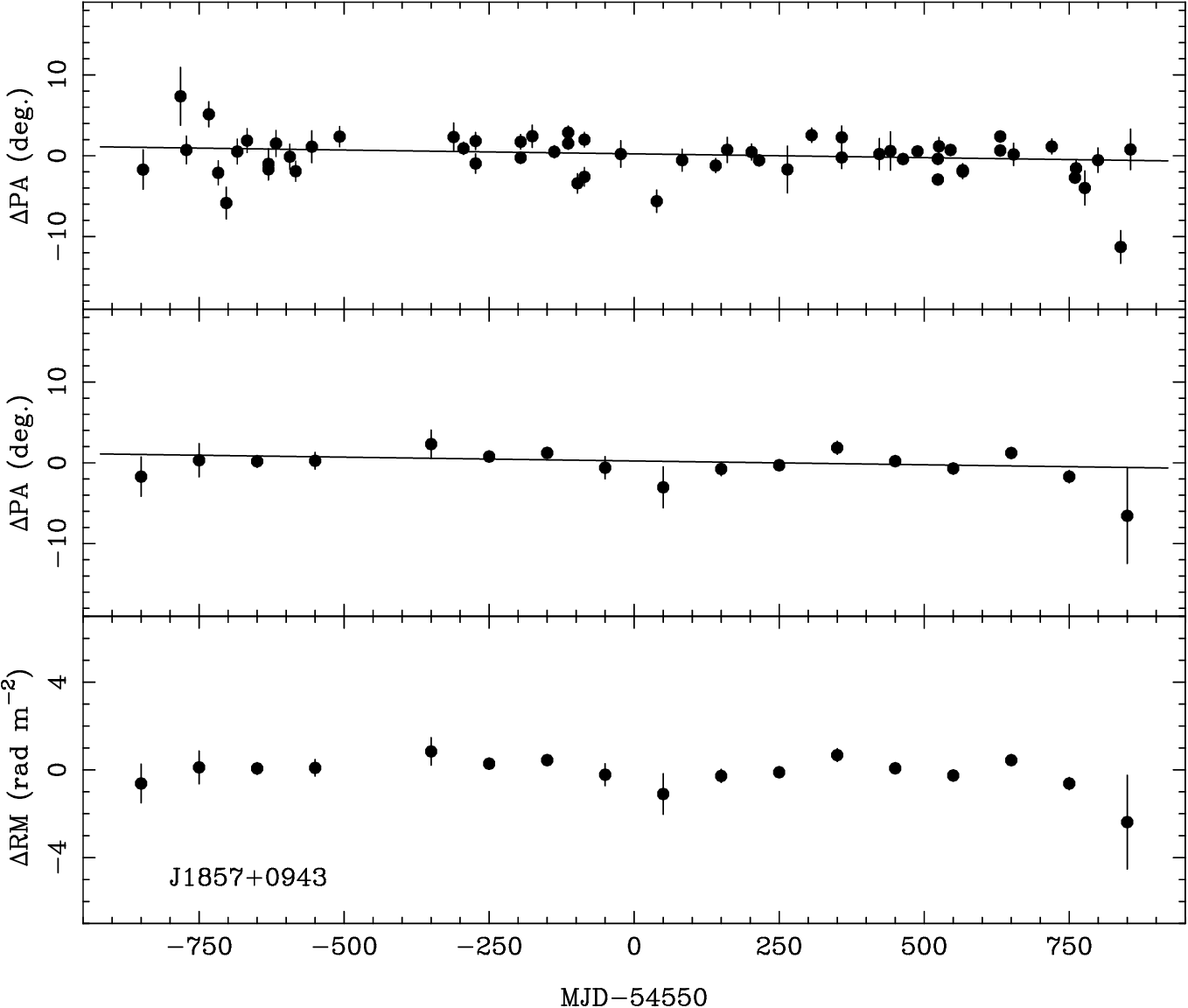}
\includegraphics[angle=-90,scale=.328]{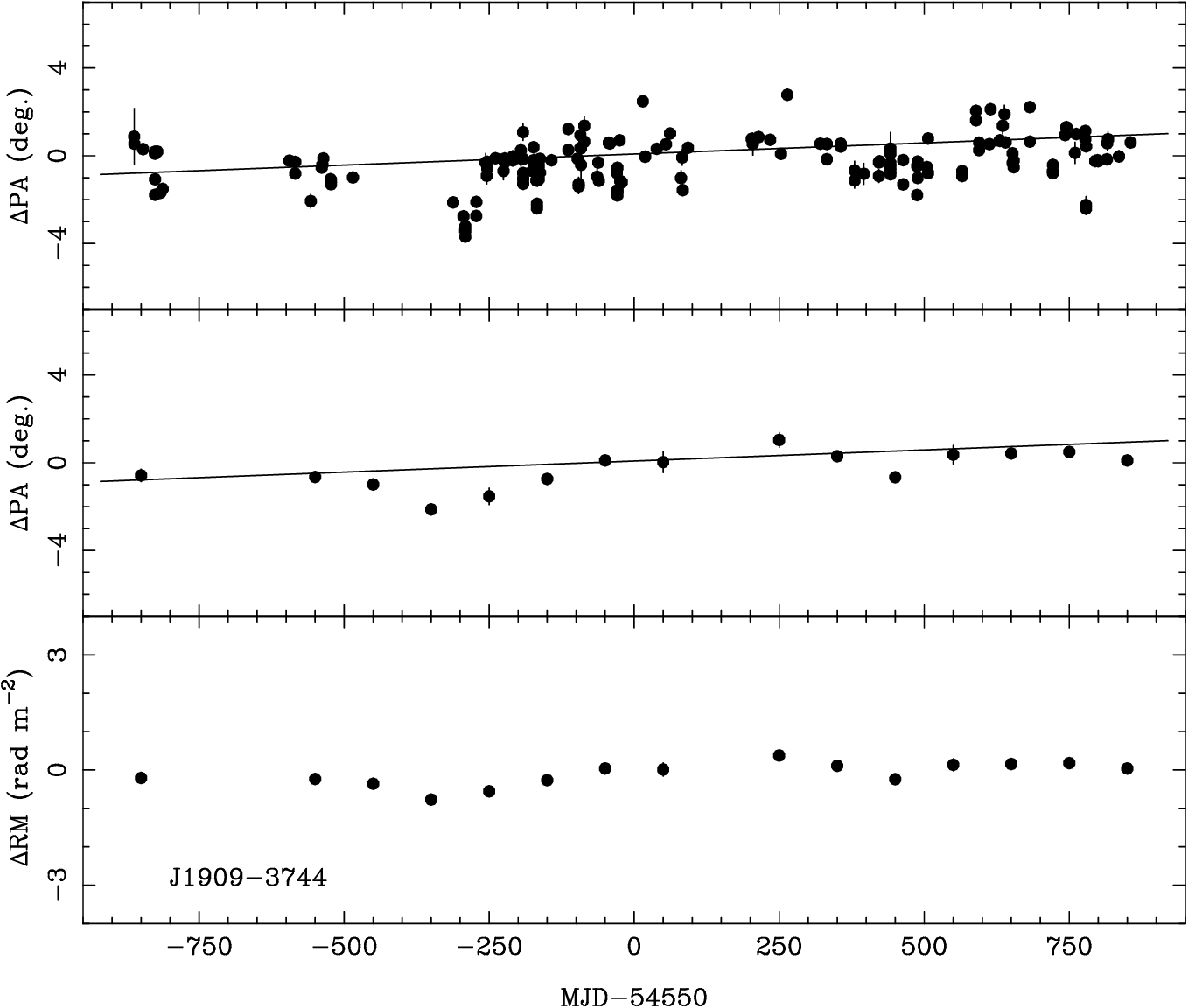}
\includegraphics[angle=-90,scale=.328]{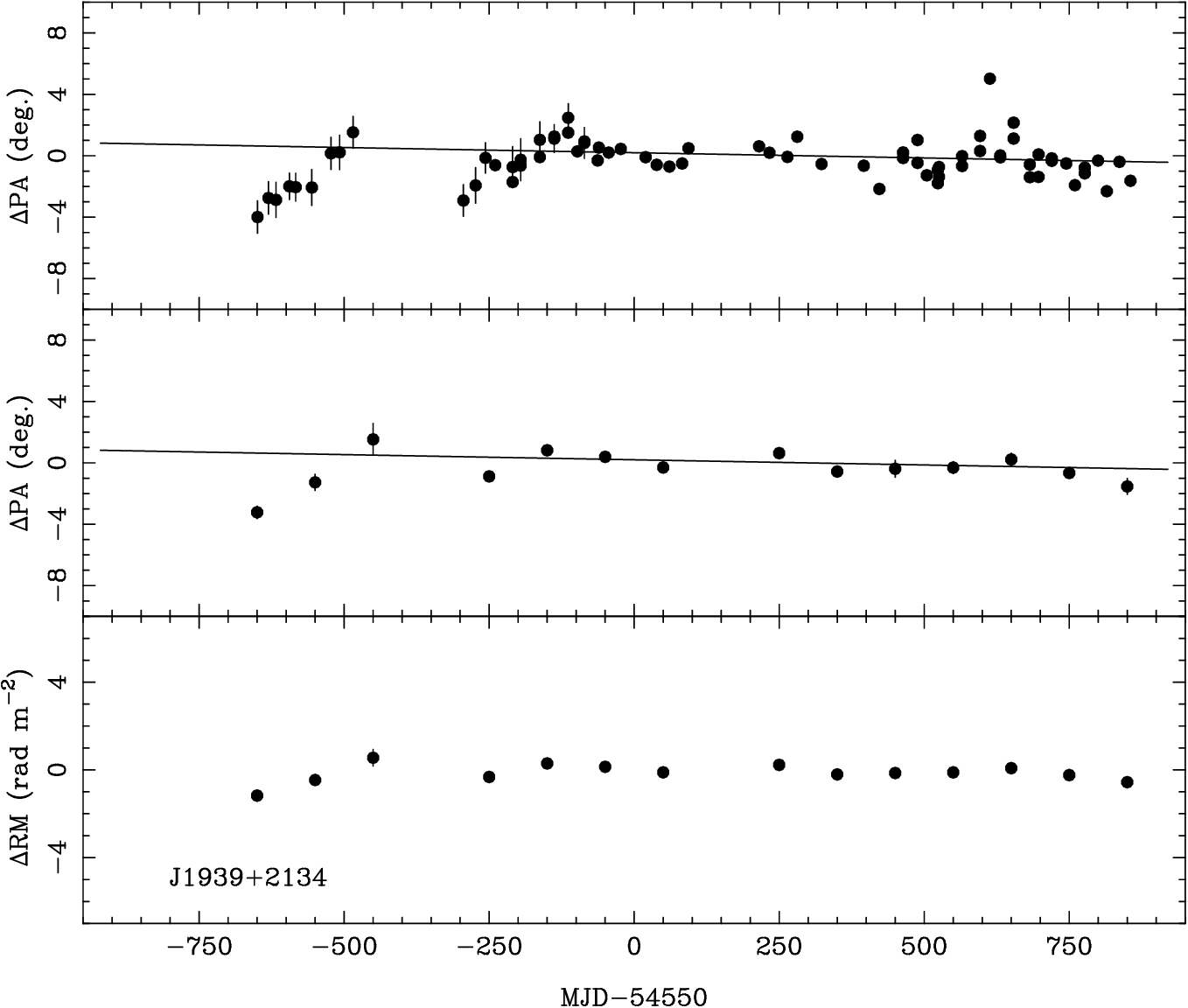}
\includegraphics[angle=-90,scale=.328]{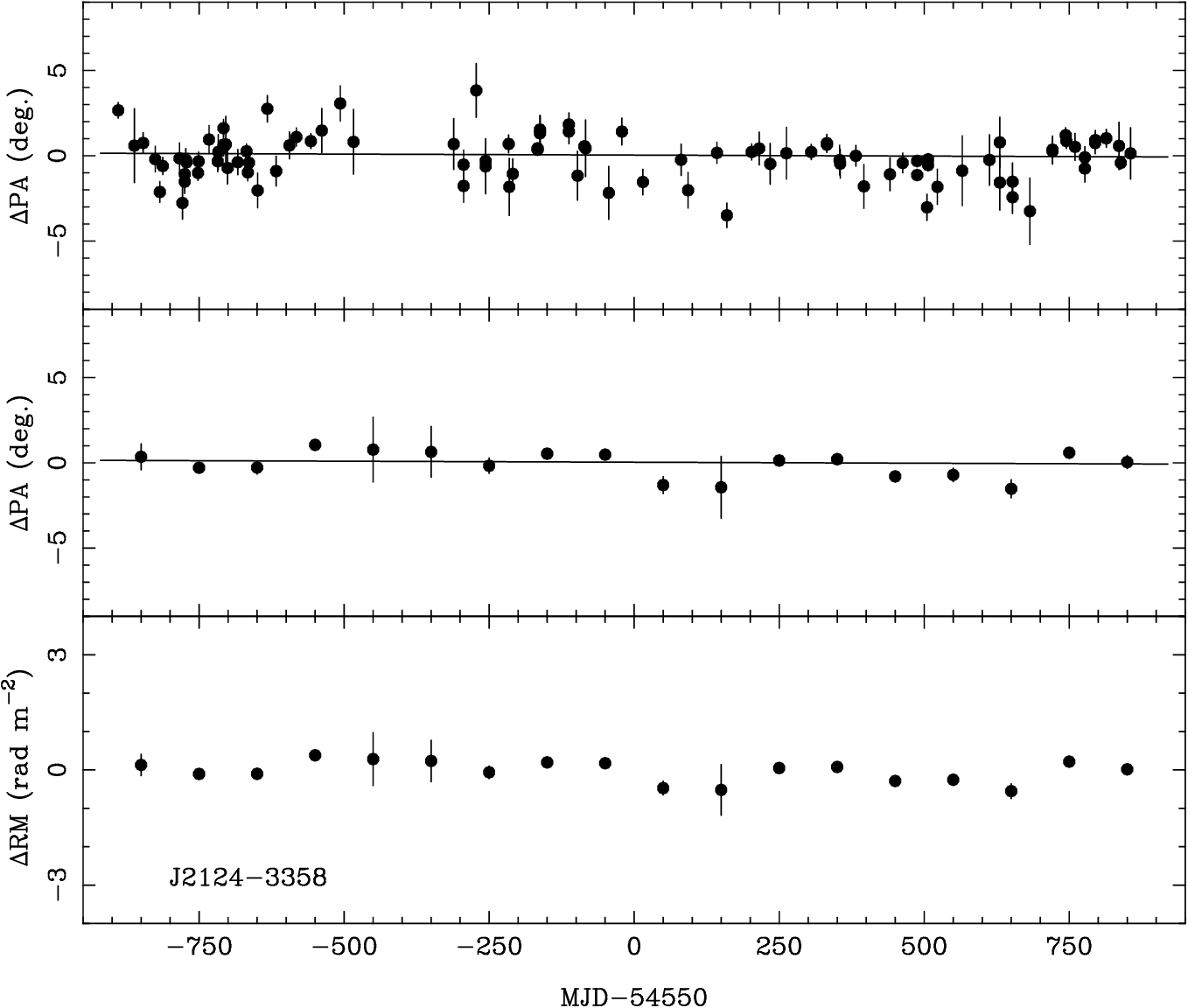}
\caption{PA variations for six of the PPTA pulsars. 
See Figure~\ref{Fig:pavar1} for
  details.\label{Fig:pavar3}}
\end{figure*}

\begin{figure*}
\centering
\centering
\includegraphics[angle=-90,scale=.328]{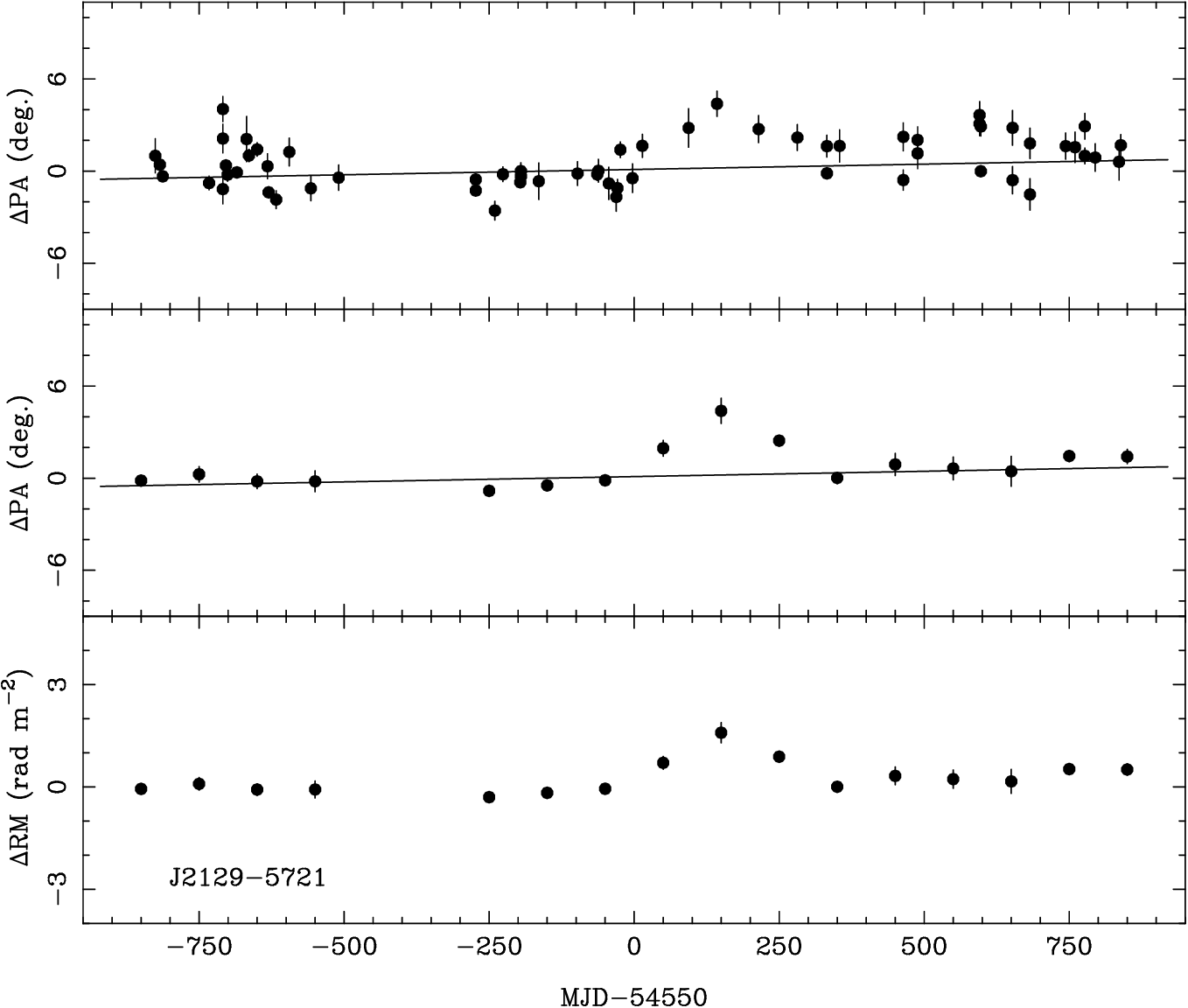}
\includegraphics[angle=-90,scale=.328]{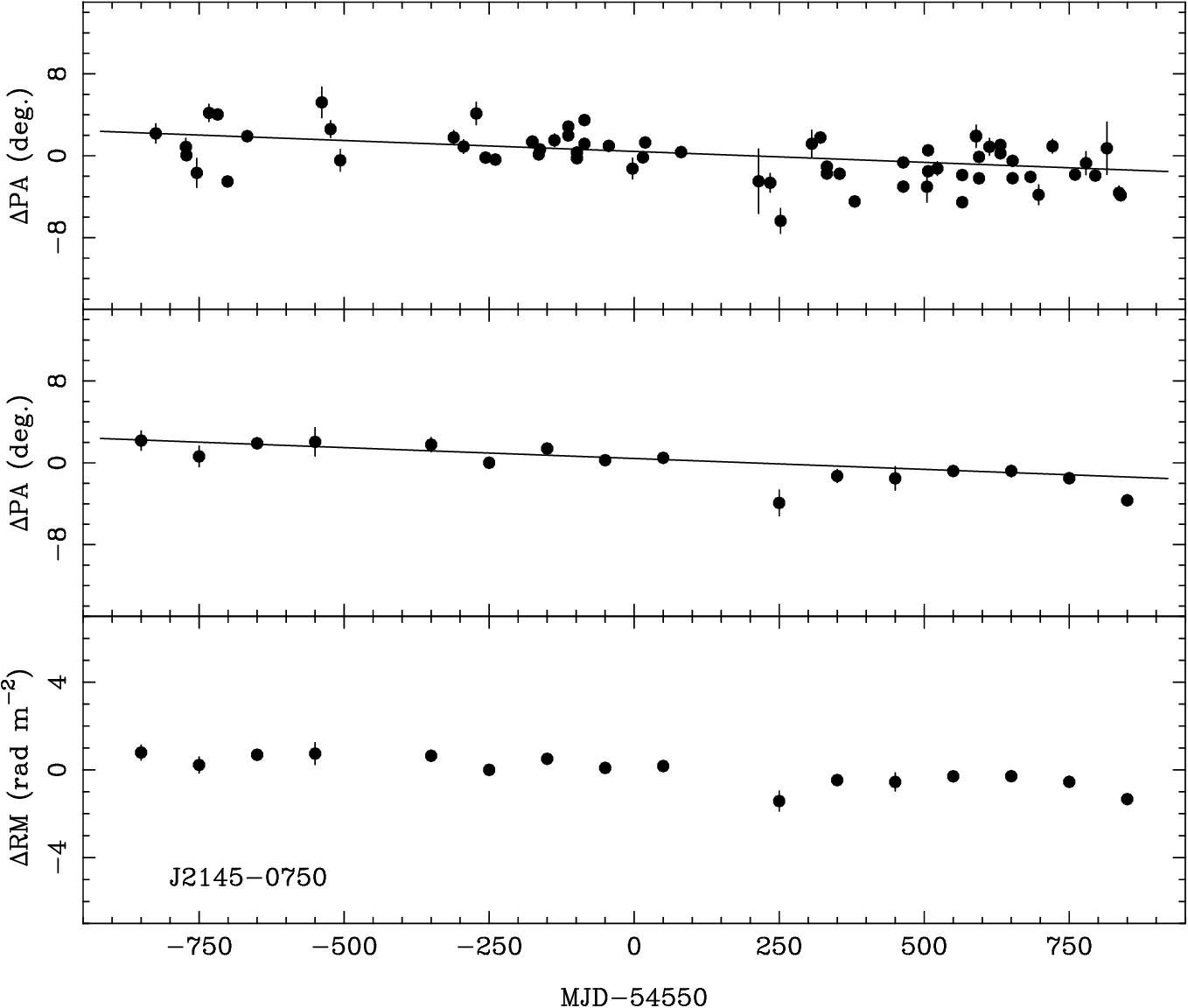}
\caption{PA variations for two of the PPTA pulsars. 
See Figure~\ref{Fig:pavar1} for
  details.\label{Fig:pavar4}}
\end{figure*}

Figures~\ref{Fig:pavar1} -- \ref{Fig:pavar4} show the IRI-corrected
PAs for the PPTA pulsars. The straight line on each plot results from
a weighted least-squares fit of a straight line to the data. To reduce
the short-term scatter, weighted averages of the corrected PAs over
100-day blocks are shown in the middle panel, again with the fitted
straight line. If we assume that there has been no variation in the
intrinsic PA over this interval, we can convert the observed PA
variations to equivalent RM variations using Equation~\ref{eq:fara} as
shown in the lower panel. Table~\ref{tb:estpar} gives the slope of the
fitted line and its estimated uncertainty and the corresponding RM
gradient for the 20 pulsars. Only a few of these gradients are
significant at more than the $3\sigma$ level; mostly the observations
are consistent with no long-term change in PA or interstellar RM
within the uncertainties.

\begin{table*}
\small
\centering
\begin{minipage}[]{120mm}

\caption[]{PA gradients, implied RM gradients, DM gradients and
  apparent line-of-sight magnetic field in the time-varying region for the
  PPTA pulsars.}\label{tb:estpar}\end{minipage}
\tabcolsep 6mm
 \begin{tabular}{ccccc}
  \hline\noalign{\smallskip}
PSR   &$d{\rm PA}/dt$   &$d{\rm RM}/dt$ & $d{\rm DM}/dt$  &$\langle B_{||}\rangle_{\rm var}$ \\
	      &($\degr$yr$^{-1}$)  &(rad m$^{-2}$ yr$^{-1}$) &(cm$^{-3}$ pc yr$^{-1}$)  &($\mu$G)\\
  \hline\noalign{\smallskip}
J0437$-$4715     &$0.08\pm 0.03$      &$0.03\pm 0.01$      &$(3\pm 3)\times10^{-5}$   &$1200\pm 1300$ \\
J0613$-$0200    &$0.64\pm 0.13$        &$0.23\pm 0.05$     &$(4.3\pm 1.6)\times10^{-4}$  &$660\pm 280$  \\
J0711$-$6830    &$-0.08\pm 0.13$      &$-0.03\pm 0.05$  &$(-5.3\pm 7.8)\times10^{-4}$  &$70\pm 160$  \\
J1022$+$1001    &$0.35\pm 0.06$   &$0.13\pm 0.02$   &$(-2.0\pm 5.5)\times10^{-4}$   &$-800\pm 2200$ \\
J1024$-$0719    &$-0.08\pm 0.09$   &$-0.03\pm 0.03$  &$(4.8\pm 1.9)\times10^{-3}$    &$-8\pm 8$ \\\\ 
J1045$-$4509     &$-0.12\pm 0.09$   &$-0.04\pm 0.03$   &$(-1.03\pm 0.01)\times10^{-2}$   &$5\pm 4$  \\
J1600$-$3053    &$0.05\pm 0.07$    &$0.02\pm 0.02$   &$(-9\pm 10)\times10^{-4}$     &$-30\pm 40$   \\
J1603$-$7202   &$0.33\pm 0.08$     &$0.12\pm 0.03$   &$(3.4\pm 0.8)\times10^{-3}$    &$40\pm 10$  \\
J1643$-$1224   &$-0.11\pm 0.08$   &$-0.04\pm 0.03$   &$(-3.2\pm 3.7)\times10^{-4}$   &$150\pm 210$  \\
J1713$+$0747       &$-0.08\pm 0.08$   &$-0.03\pm 0.03$   &$(-6\pm 24)\times10^{-5}$    &$620\pm 2500$   \\\\
J1730$-$2304    &$0.17\pm 0.12$    & $0.06\pm 0.05$   &$(-2.3\pm 7.1)\times10^{-4}$   &$-320\pm 1000$  \\
J1732$-$5049     &$0.47\pm 0.30$    &$0.17\pm 0.11$      &$(7\pm 11)\times10^{-4}$    &$300\pm 510$  \\
J1744$-$1134   &$0.04\pm 0.06$     &$0.01\pm 0.02$    &$(-5.0\pm 8.7)\times10^{-4}$    &$-20\pm 70$  \\
J1824$-$2452     &$-0.02\pm 0.10$  &$-0.01\pm 0.04$   &$(6.2\pm 0.7)\times10^{-3}$      &$-2\pm 8$  \\
J1857$+$0943   &$-0.35\pm 0.21$    &$-0.13\pm 0.08$    &$(-1.4\pm 3.3)\times10^{-3}$    &$110\pm 280$  \\\\
J1909$-$3744   &$0.37\pm 0.07$   &$0.13\pm 0.02$   &$(-4.6\pm 0.9)\times10^{-4}$    &$-350\pm 90$  \\ 
J1939$+$2134       &$-0.25\pm 0.11$  &$-0.09\pm 0.04$   &$(3.3\pm 1.7)\times10^{-4}$  &$-340\pm 230$   \\
J2124$-$3358    &$-0.04\pm 0.07$    &$-0.02\pm 0.02$   &$(-4.5\pm 8.1)\times10^{-4}$   &$50\pm 110$  \\
J2129$-$5721     &$0.25\pm 0.10$     &$0.09\pm 0.04$      &$(5\pm 4)\times10^{-4}$   &$220\pm 200$  \\
J2145$-$0750    &$-0.78\pm 0.14$     &$-0.28\pm 0.05$   &$(3.5\pm 4.6)\times10^{-4}$  &$-990\pm 1300$   \\
  \noalign{\smallskip}\hline
\end{tabular}
\end{table*}

As discussed in \S\ref{sect:intro}, we can use the ratio of the RM
gradient to the DM gradient to estimate the line-of-sight component of
the magnetic field in a putative time-varying region. DM variations
for the PPTA pulsars were measured by \citet{yhc+07}. There is only
one year of DM data overlap with the PA variation data but the DM
gradients given in Table~\ref{tb:estpar}, obtained by fitting to this
overlap year, are representative of the longer-term gradients. Because
of the high-precision timing possible for these MSPs, the absolute
changes in DM, while significant in some cases, are extremely
small. This, combined with the fact that we effectively have just
upper limits on $d{\rm DM}/dt$ for most of our sample, means that, for
these pulsars, we can just place upper limits of several $100\;\mu$G on
the interstellar field strength in the time-varying region. Values and
uncertainties of $\langle B_{||}\rangle_{\rm var}$ given in
Table~\ref{tb:estpar} have been computed using Equation~\ref{eq:bvar}
with the uncertainties in $d{\rm DM}/dt$ and $d{\rm RM}/dt$ assumed to be
independent.

Apparently significant long-term PA and implied RM variations
($>3\sigma$) are observed for five of the pulsars: PSRs J0613$-$0200,
J1022$+$1001, J1603$-$7202, J1909$-$3744 and J2145$-$0750. For PSRs
J1022$+$1001 and J2145$-$0750, $d{\rm DM}/dt$ is not well determined
and so the mean field is very uncertain. For PSRs J0613$-$0200,
J1603$-$7202 and J1909$-$3744, there are significant values for both
$d{\rm RM}/dt$ and $d{\rm DM}/dt$, so we can compute values for
$\langle B_{||}\rangle_{\rm var}$ as listed in Table~\ref{tb:estpar}. For most
of the pulsars, the $\langle B_{||}\rangle_{\rm var}$ values or limits are much
larger than typical interstellar fields of a few $\mu$G \citep{hml+06}
and even the field expected in magnetic filaments associated with
supernova remnants \citep{hhc85,rcip88}.

For PSRs J1024$-$0719, J1045$-$4509 and J1824$-$4509 the upper limits
on $\langle B_{||}\rangle_{\rm var}$ are interestingly low, of order
$10\;\mu$G. This is comparable to the estimated interstellar total
field strength, including both large-scale regular fields and
small-scale irregular fields \citep{bkb85,hfm04}. These results show
that variations in the interstellar RM for these pulsars are not
dominated by any strong filament along the line of sight.

As well as the significant long-term variations in RM discussed above,
Figures~\ref{Fig:pavar1} -- \ref{Fig:pavar4} show that there are
apparently significant fluctuations in the smoothed RM variations on a
timescale of a few hundred days for some of the MSPs. For PSR
J1939+2134, the PA appears to increase at a rate of about 3~rad
m$^{-2}$~yr$^{-1}$ for the first 150 days (MJD 53900 -- 54050). Over
this same interval, the DM is decreasing at about $-8\times
10^{-4}$~cm$^{-3}$pc~yr$^{-1}$ \citep{yhc+07}. If both these
variations are occurring in a single region, they imply a
line-of-sight magnetic field of about 4.5~mG in the region. Although
such fields are observed close to stars, such an intersection is
highly improbable (except for the Sun -- discussed below). Most
probably, the observed variations are not common to a single region,
but represent statistical fluctuations in the accumulated RM and DM
over the whole path resulting from random and uncorrelated spatial and
temporal variations in the interstellar electron density and magnetic
field.

It is important to note that the observed RM variations would be
expected to be less than the DM variations since the RM is a vector
sum (\eqref{eq:rm}) whereas the DM is a scalar sum
(\eqref{eq:dm}). Variations in regions of oppositely directed field
will cancel in RM but not in DM. A more detailed examination of the
origin of these variations would require realistic modelling of the
electron density and magnetic field fluctuations along representative
Galactic paths. This is left for future work. 

It is also worth noting that the aliased diurnal variations in ionospheric
RM are approximately covariant with annual variations in RM and DM
resulting from the changing path through the solar corona as the Earth
moves around the Sun. Both have maxima when the pulsar transits near
local noon.  At least for paths that do not pass too close to the Sun, the
DM variations are roughly sinusoidal; {\sc Tempo2} has a correction
term for delays occurring in the solar corona
\citep{ehm06,yhc+07a}. Predicted RM variations are more complex
because of structure in the solar magnetic field \citep{ojs07}. Some
of the PPTA pulsars (notably PSRs J1022+1001 and J1730$-$2304) have
low ecliptic latitudes, but the few observations when these pulsars
were close to the Sun have been excluded from the present
analysis. The fact that little or no annual PA variation is evident in
Figures~\ref{Fig:pavar1} -- \ref{Fig:pavar4} shows that any variations
due to the changing path through the solar corona are below the level
of detectabilty for these observations of the PPTA pulsars.

The main goals of the PPTA and other similar PTA projects depend on
precise measurement of pulse times of arrival (ToAs). In principle,
the ionospheric RM should be taken into account when summing the data
in frequency. Failure to do so would result in some depolarization of
the signal. Depending on the ToA algorithm used and the details of the
profile polarization this could affect the ToAs. Fortunately, at the
frequencies of 1.4 GHz and above commonly used for PTA observations,
the effect is very small, with the depolarization being $\lapp 1$\ per
cent across the band. Some ToA algorithms, for example, matrix
template matching \citep{van06}, solve for the PA rotation across the
band and so the effect is decoupled from the ToA measurement. This is
potentially important for pulsar timing at low radio frequencies.

\section{Conclusions}
\label{sect:concl}
The regular observations of 20 MSPs undertaken as part of the PPTA
project provide an excellent data base for investigation of long-term
variations in the properties of the observed pulses. In this paper we
have analysed the time variations of mean absolute pulse position angle over
4.8 years. The largest systematic effect found was that due to
variations in the Earth's ionosphere. Two ionospheric models, {\sc
  farrot} and the International Reference Ionosphere (IRI) model, were
used to predict the variations in the ionospheric component, RMi, and
to compensate the observed PA changes for ionospheric variations. It
was found that the IRI model gave better results than {\sc farrot}.

After compensation, the rms fluctuation in the PAs was about $1\degr$
with little or no significant long-term variations over the data span
for most pulsars. This implies not only that the polarization of the
emitted radiation was very stable, but also that the net long-term
variations in interstellar RM along the path to the pulsar were also
very small, typically less than 0.1 rad m$^{-2}$ yr$^{-1}$ in absolute
value. The observed variations were dominated by short-term
fluctuations that most likely originate from residual calibration
errors or errors in the RMi due to unmodelled short-timescale
variations in the ionospheric total electron content.

Apparently significant PA/RM variations on timescales of a few
100 days or longer were observed in some of the pulsars. Comparison
with the DM variations published by \citet{yhc+07} showed that these
RM variations imply high values of magnetic field if they
are interpreted as correlated variations occurring in a small part of
the path. More likely, they represent statistical fluctuations in the
whole-of-path RM resulting from random spatial and temporal variations
in both the interstellar electron density and magnetic field.

\acknowledgments This work was supported by NSFC project 10673021, the
Knowledge Innovation Programme of the Chinese Academy of Sciences,
grant no. KJCX2-YW-T09, National Basic Research Programme of China
(973 Programme 2009CB824800), the Program of the Light in China's
Western Region (LCWR) under grant XBBS201022 and the Key Laboratory of
Radio Astronomy, Chinese Academy of Sciences.  We thank  Andrew Gray
and Ken Tapping of the Dominion Radio Astrophysical Observatory for
making the Penticton ionospheric modelling program {\sc farrot} available to
us. The Parkes radio telescope is part of the Australia Telescope,
which is funded by the Commonwealth of Australia for operation as a
National Facility managed by the Commonwealth Scientific and
Industrial Research Organisation.

%\bibliographystyle{mn2e}
%\bibliography{journals,modrefs,psrrefs,crossrefs}

\begin{thebibliography}{}

\bibitem[\protect\citeauthoryear{{Beuermann}, {Kanbach} \&
  {Berkhuijsen}}{{Beuermann} et~al.}{1985}]{bkb85}
{Beuermann} K.,  {Kanbach} G.,    {Berkhuijsen} E.~M.,  1985, A\&A, 153, 17

\bibitem[\protect\citeauthoryear{{Breton}, {Kaspi}, {Kramer}, McLaughlin,
  Lyutikov, Ransom, Stairs, Ferdman \& Camilo}{{Breton} et~al.}{2008}]{bkk+08}
{Breton} R.~P.,  {Kaspi} V.~M.,  {Kramer} M.,  McLaughlin M.~A.,  Lyutikov M.,
  Ransom S.~R.,  Stairs I.~H.,  Ferdman R.~D.,    Camilo F.,  2008,
  Science, 321, 104

\bibitem[\protect\citeauthoryear{{Edwards}, {Hobbs} \& {Manchester}}{{Edwards}
  et~al.}{2006}]{ehm06}
{Edwards} R.~T.,  {Hobbs} G.~B.,    {Manchester} R.~N.,  2006, MNRAS, 372, 1549

\bibitem[\protect\citeauthoryear{{Garner}, {Gaussiran} II, {Tolman}, {Harris},
  {Calfas} \& {Gallagher}}{{Garner} et~al.}{2008}]{ggt+08}
{Garner} T.~W.,  {Gaussiran} II T.~L.,  {Tolman} B.~W.,  {Harris} R.~B.,
  {Calfas} R.~S.,    {Gallagher} H.,  2008, Adv. Space Res., 42, 720

\bibitem[\protect\citeauthoryear{{Glampedakis}, {Andersson} \&
  {Jones}}{{Glampedakis} et~al.}{2009}]{gaj09}
{Glampedakis} K.,  {Andersson} N.,    {Jones} D.~I.,  2009, MNRAS, 394, 1908

\bibitem[\protect\citeauthoryear{Hamilton, Hall \& Costa}{Hamilton
  et~al.}{1985}]{hhc85}
Hamilton P.~A.,  Hall P.~J.,    Costa M.~E.,  1985, MNRAS, 214, 5{P}

\bibitem[\protect\citeauthoryear{Hamilton, McCulloch, Manchester, Ables \&
  Komesaroff}{Hamilton et~al.}{1977}]{hmm+77}
Hamilton P.~A.,  McCulloch P.~M.,  Manchester R.~N.,  Ables J.~G.,
  Komesaroff M.~M.,  1977, Nature, 265, 224

\bibitem[\protect\citeauthoryear{Han, Ferri\`ere \& Manchester}{Han
  et~al.}{2004}]{hfm04}
Han J.-L.,  Ferri\`ere K.,    Manchester R.~N.,  2004, ApJ, 610, 820

\bibitem[\protect\citeauthoryear{Han, Manchester, Lyne, Qiao \& van
  Straten}{Han et~al.}{2006}]{hml+06}
Han J.~L.,  Manchester R.~N.,  Lyne A.~G.,  Qiao G.~J.,    van Straten W.,
  2006, ApJ, 642, 868

\bibitem[\protect\citeauthoryear{{Hobbs}, {Edwards} \& {Manchester}}{{Hobbs}
  et~al.}{2006}]{hem06}
{Hobbs} G.~B.,  {Edwards} R.~T.,    {Manchester} R.~N.,  2006, MNRAS, 369, 655

\bibitem[\protect\citeauthoryear{{Hotan}, {van Straten} \&
  {Manchester}}{{Hotan} et~al.}{2004}]{hvm04}
{Hotan} A.~W.,  {van Straten} W.,    {Manchester} R.~N.,  2004, PASA, 21, 302

\bibitem[\protect\citeauthoryear{{Johnston}, {Ball}, {Wang} \&
  {Manchester}}{{Johnston} et~al.}{2005}]{jbwm05}
{Johnston} S.,  {Ball} L.,  {Wang} N.,    {Manchester} R.~N.,  2005, MNRAS,
  358, 1069

\bibitem[\protect\citeauthoryear{Manchester}{Manchester}{1974}]{man74}
Manchester R.~N.,  1974, ApJ, 188, 637

\bibitem[\protect\citeauthoryear{Manchester}{Manchester}{2008}]{man08}
Manchester R.~N.,  2008, in {Bassa} C.,  {Wang} Z.,  {Cumming} A.,   {Kaspi}
  V.~M.,  eds, {40 Years of Pulsars: Millisecond Pulsars, Magnetars and More}
  Vol.~983, {The Parkes Pulsar Timing Array Project}.
American Institute of Physics, New York, pp 584--592

\bibitem[\protect\citeauthoryear{{Manchester}, {Kramer}, {Stairs}, {Burgay},
  {Camilo}, {Hobbs}, {Lorimer}, {Lyne}, {McLaughlin}, {McPhee}, {Possenti},
  {Reynolds} \& {van Straten}}{{Manchester} et~al.}{2010}]{mks+10}
{Manchester} R.~N.,  {Kramer} M.,  {Stairs} I.~H.,  {Burgay} M.,  {Camilo} F.,
  {Hobbs} G.~B.,  {Lorimer} D.~R.,  {Lyne} A.~G.,  {McLaughlin} M.~A.,
  {McPhee} C.~A.,  {Possenti} A.,  {Reynolds} J.~E.,    {van Straten} W.,
  2010, ApJ, 710, 1694

\bibitem[\protect\citeauthoryear{{Noutsos}, {Johnston}, {Kramer} \&
  {Karastergiou}}{{Noutsos} et~al.}{2008}]{njkk08}
{Noutsos} A.,  {Johnston} S.,  {Kramer} M.,    {Karastergiou} A.,  2008, MNRAS,
  386, 1881

\bibitem[\protect\citeauthoryear{{Ord}, {Johnston} \& {Sarkissian}}{{Ord}
  et~al.}{2007}]{ojs07}
{Ord} S.~M.,  {Johnston} S.,    {Sarkissian} J.,  2007, Solar Phys.,
245, 109

\bibitem[\protect\citeauthoryear{Rankin, Campbell, Isaacman \& Payne}{Rankin
  et~al.}{1988}]{rcip88}
Rankin J.~M.,  Campbell D.~B.,  Isaacman R.~B.,    Payne R.~R.,  1988, A\&A,
  202, 166

\bibitem[\protect\citeauthoryear{Staveley-Smith, Wilson, Bird, Disney, Ekers,
  Freeman, Haynes, Sinclair, Vaile, Webster \& Wright}{Staveley-Smith
  et~al.}{1996}]{swb+96}
Staveley-Smith L.,  Wilson W.~E.,  Bird T.~S.,  Disney M.~J.,  Ekers R.~D.,
  Freeman K.~C.,  Haynes R.~F.,  Sinclair M.~W.,  Vaile R.~A.,  Webster R.~L.,
    Wright A.~E.,  1996, PASA, 13, 243

\bibitem[\protect\citeauthoryear{van Ommen, D'Alesssandro, Hamilton \&
  McCulloch}{van Ommen et~al.}{1997}]{vdhm97}
van Ommen T.~D.,  D'Alesssandro F.~D.,  Hamilton P.~A.,    McCulloch P.~M.,
  1997, MNRAS, 287, 307

\bibitem[\protect\citeauthoryear{{van Straten}}{{van Straten}}{2004}]{vs04}
{van Straten} W.,  2004, ApJS, 152, 129

\bibitem[\protect\citeauthoryear{{van Straten}}{{van Straten}}{2006}]{van06}
{van Straten} W.,  2006, ApJ, 642, 1004

\bibitem[\protect\citeauthoryear{{van Straten}, {Manchester}, {Johnston} \&
  {Reynolds}}{{van Straten} et~al.}{2010}]{vmjr10}
{van Straten} W.,  {Manchester} R.~N.,  {Johnston} S.,    {Reynolds} J.~E.,
  2010, PASA, 27, 104

\bibitem[\protect\citeauthoryear{{Verbiest}, {Bailes}, {Bhat}, {Burke-Spolaor},
  {Champion}, {Coles}, {Hobbs}, {Hotan}, {Jenet}, {Khoo} \& et al.}{{Verbiest}
  et~al.}{2010}]{vbb+10}
{Verbiest} J.~P.~W.,  {Bailes} M.,  {Bhat} N.~D.~R.,  {Burke-Spolaor} S.,
  {Champion} D.~J.,  {Coles} W.,  {Hobbs} G.~B.,  {Hotan} A.~W.,  {Jenet} F.,
  {Khoo} J.,    et al. 2010, Class. Quantum Grav., 27, 084015

\bibitem[\protect\citeauthoryear{{Weisberg}, {Cordes}, {Kuan}, {Devine},
  {Green} \& {Backer}}{{Weisberg} et~al.}{2004}]{wck+04}
{Weisberg} J.~M.,  {Cordes} J.~M.,  {Kuan} B.,  {Devine} K.~E.,  {Green} J.~T.,
     {Backer} D.~C.,  2004, ApJS, 150, 317

\bibitem[\protect\citeauthoryear{{Yan}, {Manchester}, {van Straten},
  {Reynolds}, {Hobbs}, {Wang}, {Bailes}, {Bhat}, {Burke-Spolaor}, {Champion},
  {Coles}, {Hotan}, {Khoo}, {Oslowski}, {Sarkissian}, {Verbiest} \&
  {Yardley}}{{Yan} et~al.}{2011}]{ymv+11}
{Yan} W.,  {Manchester} R.~N.,  {van Straten} W.,  {Reynolds} J.,  {Hobbs} G.,
  {Wang} N.,  {Bailes} M.,  {Bhat} R.,  {Burke-Spolaor} S.,  {Champion} D.,
  {Coles} W.,  {Hotan} A.,  {Khoo} J.,  {Oslowski} S.,  {Sarkissian} J.,
  {Verbiest} J.,    {Yardley} D.,  2011, MNRAS, in press

\bibitem[\protect\citeauthoryear{{You}, {Hobbs}, {Coles}, {Manchester},
  {Edwards}, {Bailes}, {Sarkissian}, {Verbiest}, {van Straten}, {Hotan}, {Ord},
  {Jenet}, {Bhat} \& {Teoh}}{{You} et~al.}{2007}]{yhc+07}
{You} X.~P.,  {Hobbs} G.,  {Coles} W.~A.,  {Manchester} R.~N.,  {Edwards} R.,
  {Bailes} M.,  {Sarkissian} J.,  {Verbiest} J.~P.~W.,  {van Straten} W.,
  {Hotan} A.,  {Ord} S.,  {Jenet} F.,  {Bhat} N.~D.~R.,    {Teoh} A.,  2007,
  MNRAS, 378, 493

\bibitem[\protect\citeauthoryear{{You}, {Hobbs}, {Coles}, {Manchester} \&
  {Han}}{{You} et~al.}{2007a}]{yhc+07a}
{You} X.~P.,  {Hobbs} G.~B.,  {Coles} W.~A.,  {Manchester} R.~N.,    {Han}
  J.~L.,  2007a, ApJ, 671, 907

\end{thebibliography}

\label{lastpage}

\end{document}